\documentclass[lettersize,journal]{IEEEtran}
\usepackage{amsmath,amsfonts}
\usepackage{amssymb}
\usepackage{algorithmic}
\usepackage{array}
\usepackage[caption=false,font=normalsize,labelfont=sf,textfont=sf]{subfig}
\usepackage{textcomp}
\usepackage{stfloats}
\usepackage{url}
\usepackage{verbatim}
\usepackage{graphicx}
\usepackage{multirow}

\def\BibTeX{{\rm B\kern-.05em{\sc i\kern-.025em b}\kern-.08em
    T\kern-.1667em\lower.7ex\hbox{E}\kern-.125emX}}
\usepackage{balance}
\usepackage[normalem]{ulem}
\useunder{\uline}{\ul}{}

\usepackage{xcolor}

\begin{document}
\title{MobileMEF: Fast and Efficient Method for Real-Time Mobile Multi-Exposure Fusion}
\author{Lucas Nedel Kirsten, Zhicheng Fu, Nikhil Ambha Madhusudhana
\thanks{Lucas N. Kirsten (corresponding author, email: lucask@motorola.com) is with Motorola Mobility Comércio de Produtos Eletrônicos Ltda, Jaguariúna, SP 13918-900 BR. Zhicheng Fu and Nikhil Ambha Madhusudhana are with Lenovo Research.}}

\markboth{}%
{MobileMEF: Fast and Efficient Method for Multi-Exposure Fusion}

\maketitle

\begin{abstract}

Recent advances in camera design and imaging technology have enabled the capture of high-quality images using smartphones. However, due to the limited dynamic range of digital cameras, the quality of photographs captured in environments with highly imbalanced lighting often results in poor-quality images. To address this issue, most devices capture multi-exposure frames and then use some multi-exposure fusion method to merge those frames into a final fused image. Nevertheless, most traditional and current deep learning approaches are unsuitable for real-time applications on mobile devices due to their heavy computational and memory requirements.
We propose a new method for multi-exposure fusion based on an encoder-decoder deep learning architecture with efficient building blocks tailored for mobile devices. This efficient design makes our model capable of processing 4K resolution images in less than 2 seconds on mid-range smartphones. Our method outperforms state-of-the-art techniques regarding full-reference quality measures and computational efficiency (runtime and memory usage), making it ideal for real-time applications on hardware-constrained devices. Our code is available at: \url{https://github.com/LucasKirsten/MobileMEF}.

\end{abstract}


\section{Introduction}

Photographs captured in environments with highly imbalanced lighting often result in poor-quality images due to underexposed and overexposed regions. This issue arises from the limited dynamic range of digital cameras, which is significantly lower than in real-world scenes~\cite{fyuv}. High dynamic range (HDR) imaging techniques have been developed to address this limitation, with multi-exposure image fusion (MEF) being a prominent solution~\cite{survey_class,survey_deep}. MEF methods merge multiple images captured at different exposure levels to produce a single image that intends to retain the scene details and color fidelity~\cite{mertens}. Despite the advancements, many existing MEF methods rely on hand-crafted features or transformations, leading to robustness issues under varying conditions~\cite{benchmark}.

Traditional MEF methods, such as those based on Laplacian pyramids~\cite{mertens,fyuv}, are computationally intensive due to the multiple operations required for generating pyramid sub-images~\cite{benchmark}. This computational overhead becomes particularly problematic for hardware-constrained applications like smartphones, especially when processing high-resolution 4K images. Single-scale fusion methods~\cite{ssf_rgb} have been proposed to alleviate the computational burden, but they often produce lower-quality images due to noticeable seams and gray differences~\cite{fyuv,survey_class}. Additionally, while deep learning-based approaches have shown promise in improving MEF~\cite{survey_deep}, they often do not consider real-world deployment constraints, resulting in a trade-off between speed and quality that limits their application on mobile platforms~\cite{ifcnn,holoco,transmef,samt_mef}.

\begin{figure}[!t]
\centering
\includegraphics[width=0.45\textwidth]{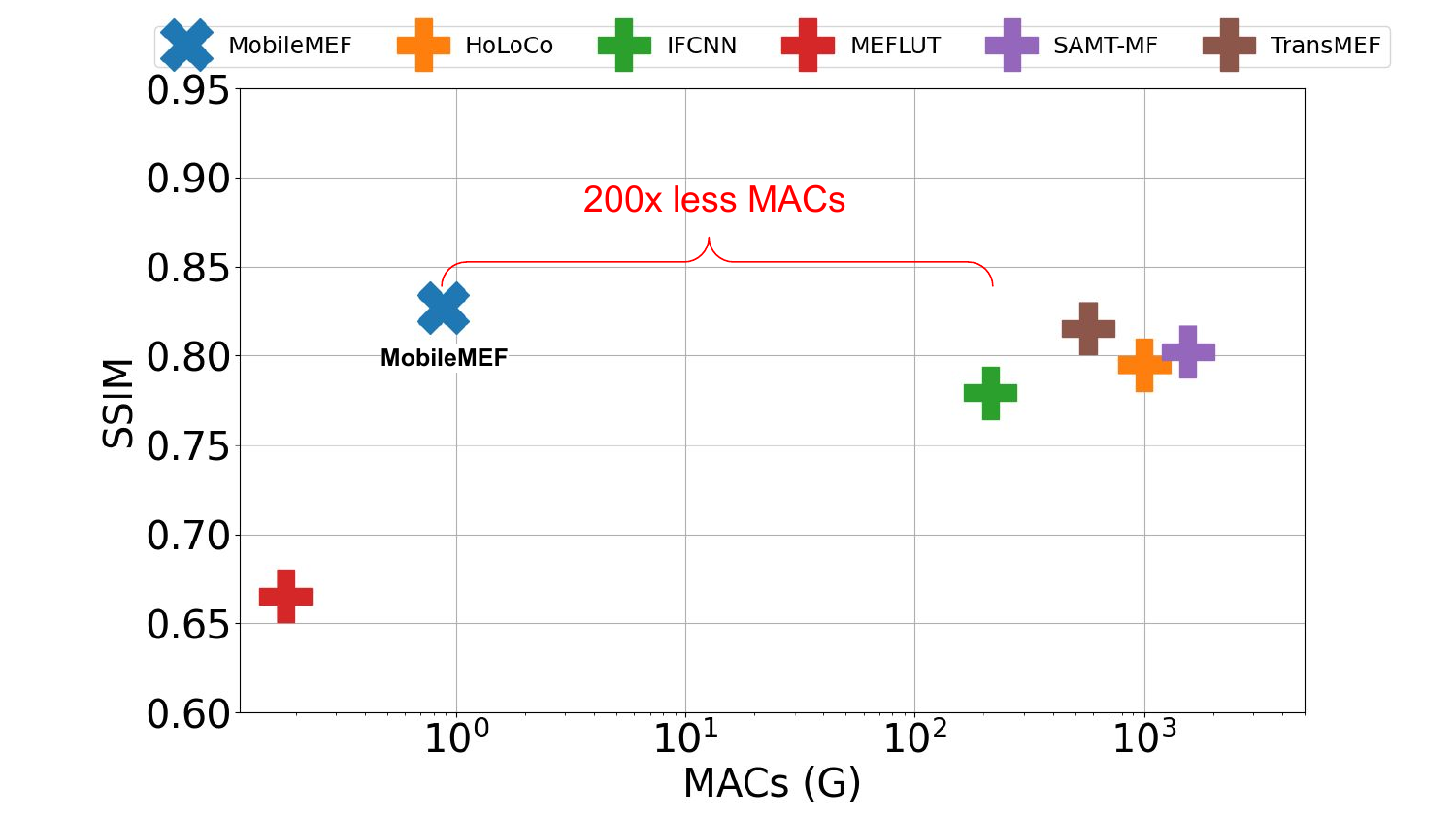}
\caption{Comparison of computational cost and performance of SOTA MEF methods: HoLoCo~\cite{holoco}, IFCNN~\cite{ifcnn}, MEFLUT~\cite{meflut}, SAMT-MEF~\cite{samt_mef}, TransMEF~\cite{transmef}. MobileMEF can process 4k images in 1.82 seconds on a mid-range mobile device using GPU.}
\label{fig:macs}
\end{figure}

To address these challenges, we propose a novel MEF method based on an Encoder-Decoder deep learning architecture designed to optimize mobile device performance, named \emph{MobileMEF}. MobileMEF draws inspiration from recent advancements in deep learning~\cite{lpienet,ghorbel2023convnext} but introduces several key modifications to enhance efficiency and effectiveness for MEF tasks. As illustrated in Fig.~\ref{fig:macs}, our method achieves superior image quality results with the lowest or near-lowest required operations compared to existing state-of-the-art (SOTA) methods. This balance of high-quality image output and low computational demand makes our method highly suitable for deployment on mobile devices, enabling efficient processing of 4K resolution images without compromising performance. Our main contributions to the field are:
\begin{itemize}
    \item We present MobileMEF, an optimized model architecture designed for hardware-constrained MEF applications. MobileMEF outperforms state-of-the-art techniques regarding full-reference quality measures and computational efficiency (runtime and memory usage), being capable of processing 4k resolution images in less than 2 seconds on mid-range smartphones;
    \item We propose a new bypass module based on Single-Scale Fusion~\cite{ssf_rgb} with YUV color space images~\cite{fyuv} that forwards an estimate of the fused image channels from the input frames to the model output predictions;
    \item We propose a new Gradient loss~\cite{ma2020structure} formulation based on cropping, capable of capturing fine details and overall image context of the predicted and ground-truth images.
\end{itemize}

\section{Related Work}

Traditional MEF methods can be divided into Multi-Scale and Single-Scale based methods~\cite{benchmark,survey_class}. Multi-Scale methods, like the Mertens~\cite{mertens} algorithm, use a sequence of Laplacian pyramids to decompose input frames into sub-images, alleviating exposure differences and smoothing local transitions by computing three quality measures related to saturation, contrast, and exposure levels. The Fast~YUV~\cite{fyuv} method improves on this by using the YUV color space and computing pyramids only on the Y channel, reducing computational efforts. In contrast, Single-Scale methods, such as the one proposed by Ancuti~et~al.~\cite{ssf_rgb}, approximate the operations in the pyramid step of the Mertens algorithm to a single step, reducing computational resource requirements.

As in many other fields, methods based on deep learning have been proposed to solve the MEF problem~\cite{survey_deep}. 
For example, the IFCNN~\cite{ifcnn} model presented a general MEF framework based on convolutional neural networks. TransMEF~\cite{transmef} is a transformer-based model that uses self-supervised multi-task learning based on an encoder-decoder architecture. HoLoCo~\cite{holoco} proposes a method based on local feature constraints built upon contrastive learning. The MEFLUT~\cite{meflut} model learns to encode the fusion weights as 1D lookup tables to achieve a highly efficient method. More recently, SAMT-MEF~\cite{samt_mef} has presented a mean teacher-based semi-supervised learning framework tailored for MEF.

In this work, we propose MobileMEF, an end-to-end encoder-decoder model designed for fast and efficient computation of MEF. Our work is built on top recent advances in deep-learning and MEF efficient designs, such as the LPIENet~\cite{lpienet} architecture, the convolutional blocks of ConvNeXt~\cite{ghorbel2023convnext}, the Fast~YUV~\cite{fyuv} and the Single-Scale MEF method proposed by Ancuti~et~al.~\cite{ssf_rgb}. Furthermore, we extend the formulation of the widely used Gradient loss~\cite{ma2020structure} to capture fine details and overall image context to improve the fused image quality.

\section{The Proposed Method}

\begin{figure*}[!t]
\centering
\includegraphics[width=0.8\textwidth]{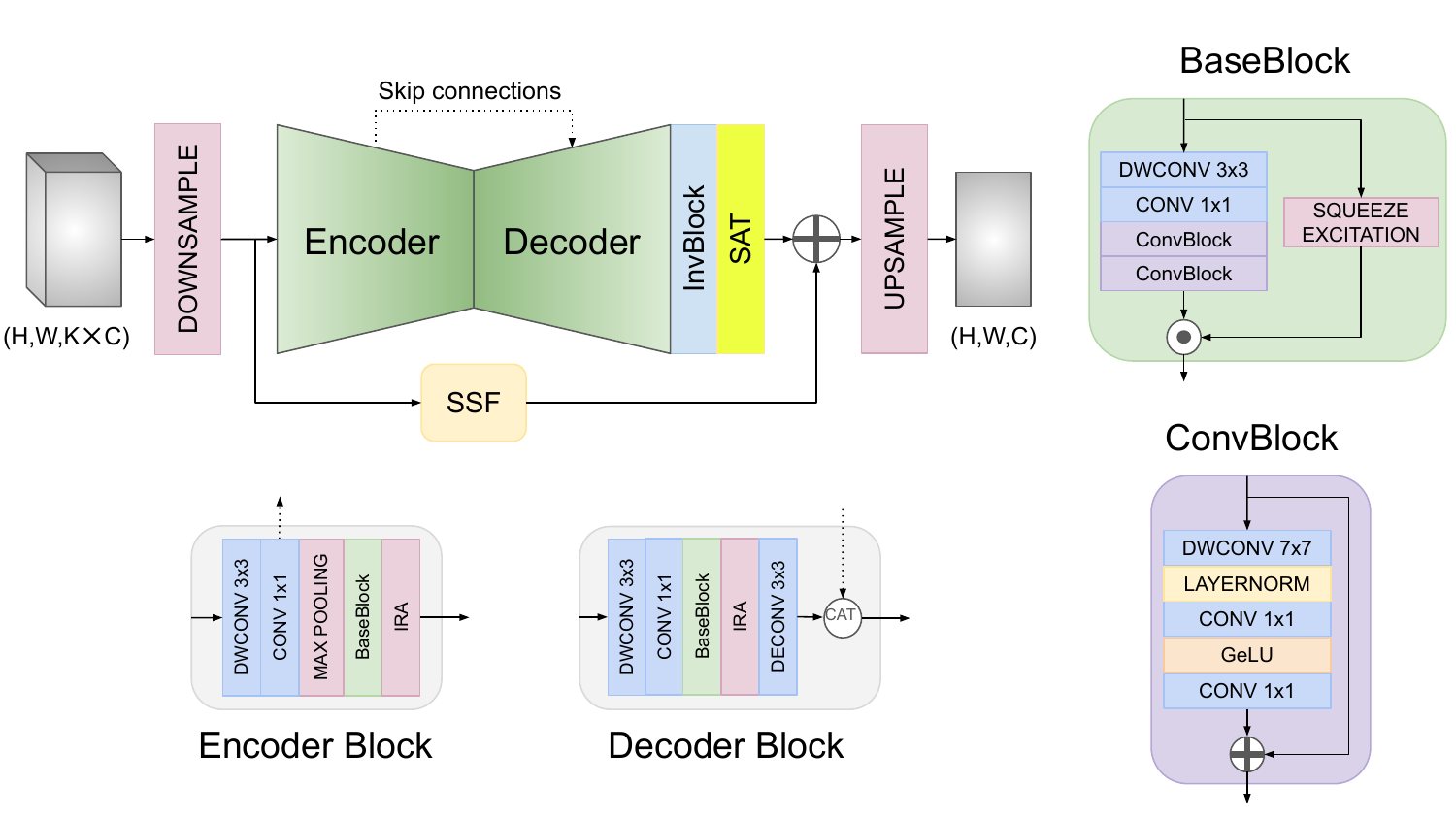}
\caption{Overall proposed MobileMEF model architecture. The encoder-decoder model receives an input with $C$ channels and $K$ frames and returns the fused frames. The SSF module merges the $K$ input frames and adds it to the model's output.}
\label{fig:model}
\end{figure*}

We propose MobileMEF, a new method for MEF based on an Encoder-Decoder deep learning architecture with efficient building blocks designed to work on mobile phones, presented in Fig.~\ref{fig:model}. The overall model architecture is inspired by the recent success of the LPIENet~\cite{lpienet} image enhancer model, but we redesigned some of their main components to work for MEF and to improve performance in mobile devices with 4K resolution images. Specifically, we redesign the model's input pipeline to work with multiple input images in the YUV color space; we replace their base convolutional blocks to use one based on ConvNeXt~\cite{ghorbel2023convnext}; added Squeeze-and-Excitation~\cite{hu2018squeeze} attention block to the Encoder and Decoder model blocks; and added a Single-Scale Fusion (SSF) bypass module based on \cite{ssf_rgb} and \cite{fyuv} that intends to forward fused information of the inputs to the model's output. We proceed to provide details regarding our method.

\subsection{Input pipeline}\label{sec:inputs}

We first concatenate the inputs' channels of the multiple input frames in the form that our model will receive inputs with shape ($H$,$W$,$K\!\times\!C$), where $H$ is the height, $W$ the width, $C$ the image channels, and $K$ the number of input frames. However, note that this simple scheme can cause computational overhead since the model requires to operate in $K$ times more channels in its first layers. 
To address this issue, we follow the proposal of other works~\cite{meflut,ifcnn} and convert the RGB input frames to the YUV color space. Converting the inputs to YUV allows us to work with smaller inputs in the UV channels since they are only related to color~\cite{fyuv}, and recent works have demonstrated that similar approaches yield superior performance with reduced computational costs (runtime and memory)~\cite{benchmark}. Hence, our input data pipeline is divided into two parts: one related to the Y channels of the input frames with shape ($H$,$W$,$K$); and the other related to the UV channels with shape ($\kappa H$,$\kappa W$,$K\!\times\! 2$), where $0<\kappa<1$ is a down-sample ratio value. Moreover, note that this scheme requires using two Encoder-Decoder models with SSF modules: one for the Y inputs, and the other for the UV, as we proceed to explain.

\subsection{Network architecture}\label{sec:network_architecture}

The overall proposed network architecture for MobileMEF is presented in Fig.~\ref{fig:model}, and is based on the success of recent works for efficient model design for mobile applications~\cite{lpienet,ghorbel2023convnext,sandler2018mobilenetv2}. For $K$ input frames with shape ($H$,$W$,$C$), where $H$ is the height, $W$ the width, and $C$ the image channels, the forward process is designed as:
\begin{enumerate}
    \item The input frames are converted to the YUV  color space and split in the channel axis. The inputs related to the UV channels are downsampled by a ratio $\kappa$. The Y and resized UV inputs are stacked in the channel axis, providing the two inputs with shape ($H$,$W$,$K$) for the Y channels, and ($\kappa H$,$\kappa W$,$K \!\times\! 2$) for the UV channels;
    \item Both inputs are downsampled by a ratio $\Gamma$;
    \item Each downsampled input is forwarded by a set of $N$ encoder and decoder blocks;
    \item The output from the decoder is then fed to a sequence composed of one Inverted residual (InvBlock)~\cite{sandler2018mobilenetv2} and one Spatial Attention (SAT) module;
    \item The downsampled inputs are processed in parallel by the SSF module that merges the $K$ input frames;
    \item The predicted fused image from the model's output is added to the merged image from the SSF module;
    \item The summed image is then upsampled by the ratio $\Gamma$ to produce the fused channel (Y or UV).
\end{enumerate}
We advocate that using a small downsampling ratio $\Gamma \leq 2$ to the input frames has negligible effects on the result fused image. However, it significantly reduces computational efforts since inputting high-resolution images to the first layers of the model can cause severe computation overheads.
We proceed to provide details regarding the blocks of the proposed network architecture following the name convention in Fig.~\ref{fig:model}.

\textbf{ConvBlock.}
We based our Convolutional Block on the ConvNeXt~\cite{ghorbel2023convnext}, as recent works have demonstrated its superiority on feature extraction~\cite{goldblum2024battle} due to its similarity with the recent Visual Transformers~\cite{dosovitskiy2020image}, but with a more efficient design.
Our convolutional blocks are composed of: a Depthwise Convolutional layer with $7\!\times\!7$ kernel size, a Layer Normalization~\cite{ba2016layer}, a Pointwise Convolution that doubles the number of input features, a GeLU~\cite{hendrycks2016gaussian} activation layer, and finally a Pointwise convolution that restores the same number of input features as the inputs. The output of the final Pointwise convolution is then added to the input of the ConvBlock.

\textbf{BaseBlock.}
Similarly with \cite{lpienet}, our Base Blocks are composed of a first $3\!\times\!3$ Depthwise Convolution layer, a Pointwise Convolution that doubles the number of input features, and a sequence of two convolutional blocks. However, we include a Squeeze and Excitation~\cite{hu2018squeeze} module that is intended to add a channel attention mechanism related to the inputs. Specifically related to the MEF problem, our experiments demonstrated that the Squeeze and Excitation module can significantly boost the results with neglectable computation overhead. We assign this performance boost to the fact that the input frames are stacked in the channel axis, where the Squeeze and Excitation module operates.

\begin{figure}[!t]
\centering
\includegraphics[width=0.4\textwidth]{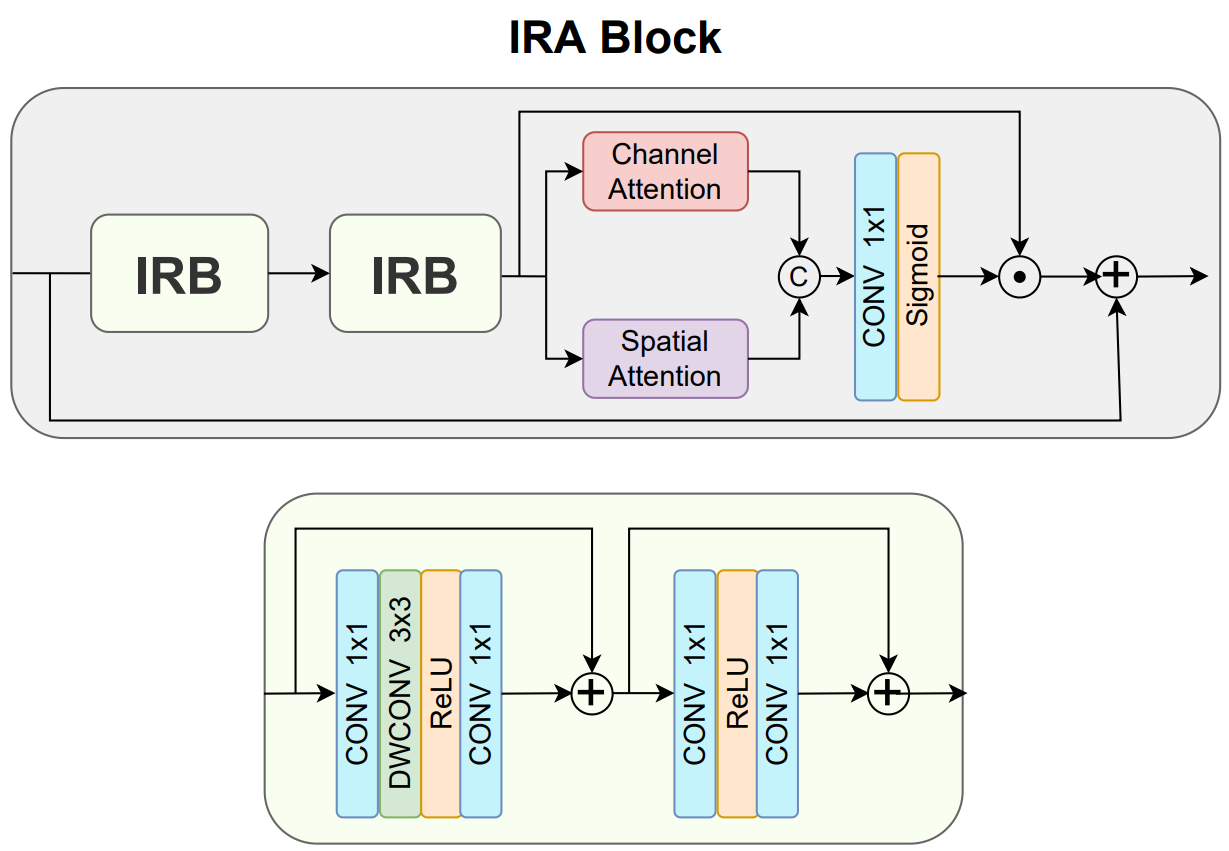}
\caption{IRA Block as proposed in LPIENet~\cite{lpienet}.}
\label{fig:ira}
\end{figure}

\textbf{Encoder Block.}
The Encoder Block compresses the inputs' spatial dimension while expanding its feature size~\cite{ronneberger2015u}. First, a $3\!\times\!3$ Depthwise Convolution layer followed by a Pointwise Convolution that doubles the number of input features is used. The output of these convolutions is stored to be used for the skip connections. Next, a Max Pooling 2D layer is employed to reduce the spatial size by a ratio of 2, and is followed by a BaseBlock and an Inverted Residual Attention (IRA) module~\cite{lpienet}. The IRA module was proposed in \cite{lpienet} to add both spatial and channel attention to each Encoder/Decoder block, and is illustrated in Fig.~\ref{fig:ira}.

\textbf{Decoder Block.}
The Decoder Block expands the spatial dimension of the input features while reducing its feature size to retain only the most important features extracted during the encoder step~\cite{ronneberger2015u}. First, we employ a $3\!\times\!3$ Depthwise Convolution layer followed by a Pointwise Convolution that halves the number of input features. Next, we use a BaseBlock and an IRA module~\cite{lpienet}. Finally, a Transposed Convolution with the same number of input and output features is used to upsample the features to be concatenated with the respective encoder layer output (i.e., the skip connection).

\textbf{InvBlock.}
The Inverted residual (InvBlock)~\cite{sandler2018mobilenetv2} module adds residual information through an efficient inverted structure, and is commonly used in many mobile applications. It comprises a branch of Pointwise convolution, Depthwise convolution, another Pointwise convolution (all with the same number of output features); and another branch composed of a Pointwise convolution with a ReLU activation. The output of both branches is then added to compose the final InvBlock output.

\textbf{SAT.}
The Spatial Attention module is designed to enhance and suppress certain spatial features of the inputs through a progressive receptive field augmentation~\cite{lpienet}. It is composed of three Convolution layers followed by ReLU (the first two) and Sigmoid activation (the last one). The convolutions use an increasing kernel size scheme of $7\!\times\!7$, $5\!\times\!5$, and $3\!\times\!3$, respectively. All convolutions have the same number of output features, that is the number of expected predicted output channels (1 for the Y channel, and 2 for the UV channels). The last output tensor (the one related to the Sigmoid activation) is then dot multiplied by the input tensor of the SAT module.

\subsection{SSF module}\label{sec:ssf_module}

Following the LPIENet~\cite{lpienet} network architecture, and as in many other image enhancement works~\cite{survey_img_enh}, we also add a skip connection that sums the input image with the network output. This simple strategy is usually used to improve model convergence and performance since it allows the network to focus only on correcting the input image, reducing the necessity of also retaining features for reconstructing it during the Encoder-Decoder step~\cite{survey_img_enh}. However, for the MEF problem, the inputs and outputs have different numbers of channels, since the main objective is to fuse the input frames into a unique output. One could argue that a simple reduction function, such as the mean or median, could be used to match the network output channels. Nevertheless, in \cite{benchmark} the authors have demonstrated that such stack techniques tend to compromise the final image quality. Moreover, they also show that Single-Scale Fusion (SSF) methods (e.g., \cite{ssf_rgb}) are capable of producing similar results to Multi-Scale ones (e.g., \cite{mertens,fyuv}), with the advantage of reducing computational efforts.

We propose to use a learnable version of the SSF method named ``SSF-YUV'' described in \cite{benchmark} for the skip input path of our model in order to provide a ``closer'' estimate of the fused input to the model prediction. We refer to it as the \emph{SSF module} and proceed to detail it. This module is based on the works of Ancuti~et~al.~\cite{ssf_rgb}, which describes a Single-Scale Fusion method based on approximations on the Mertens~\cite{mertens} algorithm; and the Fast~YUV~\cite{fyuv}, which describes an efficient MEF method designed to fuse images in the YUV color space. 

According to \cite{ssf_rgb}, a fused image channel (e.g., for RGB image, the red, green, or blue channel) can be obtained with:
\begin{equation}
\label{eq:ssf}
    \mathcal{F} = \sum_{i=1}^{K} \omega_{B} \circledast \mathcal{\overline{W}} + \alpha \cdot |\mathcal{P}_1 (\mathcal{I}_i) \odot \mathcal{P}_1 (\mathcal{\overline{W}})| ,
\end{equation}
where $K$ is the number of input frames, $\omega_{B}$ is a blur kernel, $\mathcal{\overline{W}} \in \mathbb{R}^2$ is the normalized \emph{weight maps}\footnote{Note that the term ``weights'' here have a different meaning than the one used for the weights of a deep learning model. In traditional MEF methods (such as \cite{ssf_rgb,fyuv,mertens}), the ``weight maps'' are the output of hand-coded functions applied to the input frames to extract some visual information of it, such as contrast, saturation, and exposure.} of the fusion method, $\alpha$ is a scalar constant set empirically, $\mathcal{I}_i \in \mathbb{R}^2$ is the $i$-th input frame channel, and ${\mathcal{P}_1 (I) = I - \texttt{UP}(\texttt{DOWN}(I))}$ is a single step Laplacian Pyramid function,
where \texttt{UP} and \texttt{DOWN} are up-sampling and down-sampling operations, respectively that halve the spatial dimensions. The definition of these up- and down-sampling operations is usually accompanied by convolving the inputs with Gaussian kernels. However, in our experiments, we noted that a standard resizing with bi-linear interpolation could achieve similar or superior results.

For YUV images, the UV channels relate solely to color, so using some weighted average method such as Eq.~\ref{eq:ssf} for fusing these channels would shift all colors to gray~\cite{fyuv}. To address this, we use Eq.~\ref{eq:ssf} to compute the fusion solely in the Y channel, whereas for the UV channels we use \cite{fyuv} proposal:
\begin{equation}
    \mathcal{F}_U = \underset{K}{\text{max}}~\mathcal{I}_U
\end{equation}
and
\begin{equation}
    \mathcal{F}_V = \underset{K}{\text{max}}~\mathcal{I}_V,
\end{equation}
where $\mathcal{I}_U$ and $\mathcal{I}_V$ are the U and V channels of the input frames, respectively. This formulation reasons that, when imaging, for a reasonably exposed area, the color should be bright and so its color components should be close to its maximum value.

Based on the \cite{fyuv} work, we also use two weight maps related to computing the contrast and exposure of the YUV input frames. We compute the contrast weight as:
\begin{equation}
\label{eq:constrast_w}
    \mathcal{W}_C = \omega_C \circledast \mathcal{I}_Y ,
\end{equation}
where $\omega_C$ is the Laplacian kernel, and $\mathcal{I}_Y$ is the Y channel of the input frames. And for the exposure weight:
\begin{equation}
    \mathcal{W}_E = |\mathcal{I}_U| \odot |\mathcal{I}_V| .
\end{equation}
The final normalized weight maps is computed as:
\begin{equation}
    \mathcal{\overline{W}} = \frac{\mathcal{W}_C \odot \mathcal{W}_E} 
    {\sum_{i=1}^{K} \mathcal{W}_C \odot \mathcal{W}_E}.
\end{equation}

\subsection{Loss function}\label{sec:loss_function}

We define the loss function to train our complete deep-learning model based on a weighted sum of $\mathcal{L}_1$ (mean absolute error) and Gradient loss~\cite{ma2020structure}. The final loss function is:
\begin{equation}
\label{eq:loss}
    \mathcal{L} = \mathcal{L}_1 + \lambda \cdot \mathcal{L}_{\text{Grad}},
\end{equation}
where $\mathcal{L}_{\text{Grad}}$ is the Gradient loss, and $\lambda$ is a scalar constant used to scale the loss.

For the Gradient loss, the features extracted from a large pre-trained classification model (e.g., VGG16 or VGG19~\cite{simonyan2014very}) are used to compute the $\mathcal{L}_1$ distance between the ground-truth and predicted features, to preserve image structure and generate perceptual-pleasant details. These classification models usually are trained with small inputs (e.g., $224\!\times\!244$). However, our model is intended to work with 4K high-resolution images, and feeding such large images would harm the classifier's capabilities of extracting features. One solution would be to resize the ground truth and predicted images to match the classifier input size, but it can cause the loss of fine details. Current works~\cite{lpienet,holoco} usually rely on cropping the high-resolution inputs to fit a ``more suitable'' image size (e.g., $512\!\times\!512$) to reduce the loss of fine details to the Gradient loss. Nevertheless, this is also not ideal, because it may lose some contextual information of the whole image.

We propose a simple method to improve the Gradient loss performance, based on resizing and cropping the input and ground-truth images to preserve contextual information and finer details. Given the predicted and ground-truth images ${\{\mathcal{I}(x,y),~ \mathcal{I}_\text{GT}(x,y)\} \in \mathbb{R}^{H\times W}}$ respectively (omitted the channel axis for simplicity), and a feature extractor classifier $\Psi$ trained with input size $h\!\times\!w$, we extract the crops from the images in the form:
\begin{equation}
\label{eq:crop}
\delta(\mathcal{I}) = 
    \begin{cases}
        I(x+\Delta x,y+\Delta y) & \text{if } 0 \leq x \leq w,~0 \leq y \leq h \\
    0 & \text{otherwise}
    \end{cases} ,
\end{equation}
where $\Delta x$ and $\Delta y$ are used to define the starting point for the crop region. The final Gradient loss is then computed using the following composition:
\begin{multline}
\label{eq:grad_loss}
    \mathcal{L}_{Grad} = \mathcal{L}_1(\Psi(\mathcal{I}^r), \Psi(\mathcal{I}^r_\text{GT})) + \\
    \frac{1}{M} \cdot \sum_{i=1}^{M}
    \mathcal{L}_1(\Psi(\delta_i(\mathcal{I})), \Psi(\delta_i(\mathcal{I}_\text{GT}))),
\end{multline}
where ${\{\mathcal{I}^r,~\mathcal{I}^r_\text{GT}\} \in \mathbb{R}^{h\times w}}$ are the resized input and ground-truth images respectively, and $M$ is the number of extracted crops. We argue that this definition of the Gradient loss can extract fine details and preserve image context, as we proceed to demonstrate in our results.

\section{Experimental setup}

\subsection{Dataset}\label{sec:dataset}

We employ the widely used open-source benchmark MEF dataset SICE~\cite{cai2018learning} to train and evaluate our proposed method.  It is composed of both indoor and outdoor scenes, captured with seven types of consumer-grade cameras.
We used the most common split of SICE composed of 360 scenes, of which 302 are used for training and 58 for testing. The resolution of most images is between $3000\!\times\!2000$ and $6000\!\times\!4000$ pixels. Each scene comprises at least three images of different Exposure Values (EV) with values -1, 0, and +1, plus one ground-truth image.

\subsection{Implementation details}

We employ a reduction factor of ${\kappa=1/4}$ on the UV channels (recall Sec.~\ref{sec:inputs}), and of ${\Gamma=2}$ for the input frames (both Y and UV, recall Sec.~\ref{sec:network_architecture}). In all resize operations we use the bi-linear interpolation. We use ${N=5}$ encoder/decoder blocks for the Y inputs, and ${N=3}$ for the UV inputs, since they are 4 times smaller than the Y ones. The number of features for the first Pointwise convolution of the encoder branch is set to 4 for the Y inputs, and 8 for the UV inputs. This value is doubled for each new encoder block and halved on the respective decoder one (recall Sec.~\ref{sec:network_architecture}).

For the SSF module (Eq.~\ref{eq:ssf}), we use $\alpha=0.2$ as in \cite{ssf_rgb}, and ${\omega_{B} \in \mathbb{R}^{5\times5}}$ kernel is initialized with all its values equal $1/25$. For the contrast weight (Eq.~\ref{eq:constrast_w}), ${\omega_C \in \mathbb{R}^{3\times3}}$ is initialized with values:
\begin{equation}
    \omega_C = 
    \begin{bmatrix} 
    \centering
        0 & 1 & 0 \\
        1 & -4 & 1 \\
        0 & 1 & 0
    \end{bmatrix},
\end{equation}
which is the Laplacian kernel. Both $\omega_{B}$ and $\omega_C$ are treated as learnable parameters during the model training and are implemented using Depthwise convolutions.

For the loss function, we set $\lambda=1$ since we use $\mathcal{L}_1$ for computing both $\mathcal{L}_\text{Grad}$ (Eq.~\ref{eq:grad_loss}) and the complete $\mathcal{L}$ loss (Eq.~\ref{eq:loss}). For the proposed crop-like $\mathcal{L}_\text{Grad}$ (Eq.~\ref{eq:grad_loss}), we extract $M=5$ crops of the input and ground truth images using Eq.~\ref{eq:crop}, with $(\Delta x, \Delta y) = \{{(0,0)}, {(W-w,0)}$, ${(0,H-h)}$, ${(W-w,H-h)}$, ${(\frac{W-w}{2},\frac{H-h}{2})\}}$, which correspond to the top-left, top-right, bottom-left, bottom-right and center regions of the images, respectively. The VGG19~\cite{simonyan2014very} is used as the feature extractor $\Psi$ (Eq.~\ref{eq:grad_loss}), with input size $(h,w)=(448,448)$.

The model was implemented in Python~3 programming language~\cite{python3} and using the Tensorflow~2~\cite{abadi2016tensorflow} and Keras~2~\cite{chollet2015keras} frameworks. We trained the model using the Adam optimizer~\cite{kingma2014adam} with $\beta_1=0.9$, $\beta_2=0.999$, and a fixed learning rate of $10^{-4}$. We use full-sized (i.e., no crops) images and resize them to $4096\!\times\!2816$ pixels, to allow the network to learn high-resolution features, and use batch size equals 1.
We selected two different input pipelines regarding the choice of $K=2$ frames from each scene to test the robustness of our model: (i) using the EVs -1 and +1 frames, as in \cite{meflut,ifcnn}; (ii) using the most under and over exposed frames for each scene (the EV value changes depending on the scene), as in \cite{holoco,transmef,samt_mef}. We report results for 300 training epochs, but we also show in Sec.~\ref{sec:ablation} that it can be diminished to 100 without major performance loss. The model was trained in a \texttt{NVIDIA A100-SXM4} GPU with 40GB of memory, \texttt{AMD EPYC 7J13 64-Core Processor} CPU with 30 processing units, and 200GB of RAM.

\subsection{Evaluation protocol}

We evaluated our proposed method with 11 quantitative image quality metrics, and with runtime and memory usage measurements during model inference. Our results were compared with three traditional  MEF methods\footnote{We used the implementations available at: \url{https://github.com/LucasKirsten/Benchmark-Image-Fusion}.}: Mertens~\cite{mertens}, Fast~YUV~\cite{fyuv}, and Ancuti~et~al.~\cite{ssf_rgb}; and five SOTA methods\footnote{We used the implementations provided in their official Github repositories.} based on deep learning, namely: HoLoCo~\cite{holoco}, MEFLUT~\cite{meflut}, TransMEF~\cite{transmef}, IFCNN~\cite{ifcnn}, and SAMT-MEF~\cite{samt_mef}. We proceed to provide details regarding our evaluation methodology.

\textbf{Quantitative evaluation.}
For evaluating the quality of the fused images, we employed nine full-reference metrics: Structural Similarity Index (SSIM)~\cite{ssim}, Multi-scale Structural Similarity Index (MS-SSIM)~\cite{ssim}, Peak signal-to-noise ratio (PSNR), $\Delta$E~CIEDE2000 ($\Delta$E~2000)~\cite{deltae2000}, Visual Information Fidelity (VIFp)~\cite{vifp}, Feature Similarity Index Measure (FSIM)~\cite{fsim}, Spectral Residual based Similarity (SR-SIM)~\cite{srsim}, Visual Saliency-induced Index (VSI)~\cite{vsi}, and Mean Deviation Similarity Index (MDSI)~\cite{mdsi}; and two no-reference ones: MEF-SSIM~\cite{mefssim} and Qc~\cite{qc}.

To ensure a fair comparison, as the SICE dataset includes images of varying resolutions (see Sec.~\ref{sec:dataset}), the predicted images from the tested MEF methods are resized to match the resolution of their ground-truth counterparts before computing the metrics.

\textbf{Computational resources evaluation.}\label{sec:computation_eval}
In order to establish a fair benchmark among our method and competitors, we converted all models to the ONNX~13~\cite{onnx} format, which is a flexible and popular format for distributing deep learning models for inference purposes. Then, we used the converted models for inference with dummy inputs and measured runtime and memory usage on both CPU and GPU processors. The reported measurements are the average of 20 runs for each evaluation. We tested the models in a notebook with \texttt{Intel(R) Xeon(R) W-11955M 2.60Hz} CPU with 64GB of RAM and x64-based processor, and \texttt{NVIDIA RTX A3000 Laptop} GPU with 6GB of memory. The models were evaluated with input resolution of $512^2$, $768^2$, $1024^2$, $1280^2$, $1536^2$, $1792^2$, $2048^2$, and $4096^2$. However, some methods could not be evaluated on high resolutions due to hardware limitations (i.e., out of memory error).

For the mobile evaluation, we accessed the runtime and memory usage of MobileMEF using a smartphone with 4 GB of RAM, four \texttt{2.4 GHz Kryo 265 Gold} and four \texttt{1.9 GHz Kryo 265 Silver} processors, \texttt{Snapdragon 680 4G Qualcomm SM6225} chipset, and \texttt{Adreno 610} GPU. 
Our trained model was converted to the TensorFlow Lite format\footnote{Available at: \url{https://www.tensorflow.org/lite/}} to run on Android devices. The measurements were performed with the TensorFlow Lite benchmark tool\footnote{Available at: \url{https://www.tensorflow.org/lite/performance/measurement}} with 10 simulated runs using GPU, with a fixed image resolution of $4096\!\times\!4096$ pixels.

\section{Results and discussions}

\subsection{Comparison with other works}

We present the results for our quantitative evaluation in Tab.~\ref{tab:comp_results}, supported by visual results in Figs.~\ref{fig:qualitative}~and~\ref{fig:qualitative2}. \textbf{MobileMEF achieves SOTA results for most of the full-reference metrics}: 6/9 using EVs -1 and +1 as input frames (second best in MS-SSIM and VSI), and 7/9 using the most under and over-exposed frames as input (second best in VIFp), suppressing the performance of methods that require 200 to 1700 times more operations (MACs column).
These results hint the effectiveness of our method in producing high-quality fused images with minimal computational resources, suggesting that it is well-suited for applications requiring high-quality images without excessive computational costs, such as in smartphone cameras.

Although our method does not achieve the highest MEF-SSIM and Qc values, it still performs competitively. Nevertheless, it's important to note that \emph{some SOTA methods incorporate one (or more) of the tested metrics in their loss function, potentially giving them an advantage in these metrics}. Moreover, no-reference metrics may hinder some intricacies over the predicted images compared to their ground-truth counterparts, such as penalizing regions that require the model to interpolate some content due to a lack of information in the input frames.

\begin{table*}[!t]
\centering
\caption{Quantitative results comparing MobileMEF to Traditional and deep-learning based SOTA methods using two different EV frames input pipelines. The MACs (G) column was computed using an image resolution of $1280^2$. \textbf{Bold} values mark the best results, and {\ul underline} mark the second best ones.}
\resizebox{0.98\textwidth}{!}{
\begin{tabular}{l|c|ccccccccccccc}
\textbf{Method} & \textbf{Inputs} & \textbf{MACs (G)} ↓ & \textbf{SSIM} ↑ & \textbf{MS-SSIM} ↑ & \textbf{PSNR} ↑ & \textbf{MEF-SSIM} ↑ & \textbf{Qc} ↑ & \textbf{$\Delta$E 2000} ↓ & \textbf{VIFp} ↑ & \textbf{F-SIM} ↑ & \textbf{SR-SIM} ↑ & \textbf{VSI} ↑ & \textbf{MDSI} ↓ \\ \hline

Mertens & {\multirow{9}{*}{\rotatebox{90}{EVs -1 and 1}}} & ~~~~{\ul 0.14} & 0.7327~ & 0.8679 & 16.7615~ & 0.9358~ & 0.8619 & 13.4556 & 0.3955 & 0.8982 & 0.9332 & 0.9592 & 0.3646 \\
Fast~YUV && ~~~~\textbf{0.08} & 0.7299~ & 0.8611 & 16.4935~ & 0.9060~ & 0.8549 & 14.8283 & 0.3839 & 0.8926 & 0.9314 & 0.9552 & 0.3756 \\
Ancuti~et~al. && ~~~~0.25 & 0.7543~ & 0.8834 & 17.5773~ & {\ul 0.9607}~ & \textbf{0.8927} & 12.8665 & {\ul 0.4149} & 0.9236 & 0.9460 & 0.9682 & 0.3403 \\

HoLoCo && 1004.41 & {\ul 0.7973}~ & \textbf{0.8972} & {\ul 19.9179}~ & 0.8890~ & 0.8135 & 9.8006 & 0.3801 & {\ul 0.9372} & {\ul 0.9645} & \textbf{0.9784} & {\ul 0.2997} \\
MEFLUT && ~~~~0.18 & 0.6651~ & 0.8323 & 14.5945~ & 0.9490$^\dagger$ & 0.8456 & 16.6180 & 0.3256 & 0.8859 & 0.9137 & 0.9546 & 0.3842\\
TransMEF && ~569.47 & 0.7461$^\dagger$ & 0.8807 & 17.3411~ & \textbf{0.9757}~ & {\ul 0.8872} & 12.8441 & 0.4118 & 0.9224 & 0.9447 & 0.9692 & 0.3370 \\
IFCNN && ~213.18 & 0.7794~ & 0.8697 & 19.0874$^\dagger$ & 0.9440~ & 0.7885 & {\ul 9.6578} & \textbf{0.4672} & 0.9163 & 0.9468 & 0.9719 & 0.3112\\
SAMT-MEF && 1552.58 & 0.7489$^\dagger$ & 0.8769 & 18.4657$^\dagger$ & 0.9103~ & 0.8458 & 12.0584 & 0.3339 & 0.9249 & 0.9511 & 0.9712 & 0.3215\\
\textbf{MobileMEF} && ~~~~0.88 & \textbf{0.8270}~ & {\ul 0.8937} & \textbf{20.6532}~ & 0.8965~ & 0.8010 & \textbf{9.3952} & 0.3676 & \textbf{0.9376} & \textbf{0.9647} & {\ul 0.9778} & \textbf{0.2990} \\ 

\hline

Mertens & {\multirow{9}{*}{{\rotatebox{90}{\begin{tabular}[c]{@{}c@{}}Most under and over \\ exposed frames\end{tabular}}}}} & ~~~~{\ul 0.14} & 0.7838~ & 0.8423 & 18.6549~ & 0.9030~ & 0.6602 & 10.5726 & 0.3064 & 0.8781 & 0.9158 & 0.9561 & 0.3738 \\
Fast~YUV && ~~~~\textbf{0.08} & 0.7732~ & 0.8213 & 17.9362~ & 0.7865~ & 0.6408 & 14.1170 & 0.2930 & 0.8601 & 0.9108 & 0.9458 & 0.3919 \\
Ancuti~et~al. && ~~~~0.25 & 0.7795~ & 0.8197 & 18.0990~ & 0.9048~ & {\ul 0.6696} & 10.9890 & 0.2838 & 0.8489 & 0.8858 & 0.9486 & 0.4004 \\

HoLoCo && 1004.41 & 0.7951~ & 0.8594 & 19.3602$^\dagger$ & 0.8875$^\dagger$ & 0.6551 & 9.4392 & 0.2880 & {\ul 0.9180} & {\ul 0.9516} & {\ul 0.9744} & {\ul 0.3190}\\
MEFLUT && ~~~~0.18 & 0.5454~ & 0.7264 & 10.3333~ & 0.9272$^\dagger$ & 0.4878 & 25.9636 & 0.1703 & 0.7833 & 0.8312 & 0.9244 & 0.4365\\
TransMEF && ~569.47 & \textbf{0.8156}$^\dagger$ & 0.8357 & 18.6669~ & \textbf{0.9668}~ & 0.6497 & 9.8191 & 0.2493 & 0.8778 & 0.9134 & 0.9584 & 0.3731 \\
IFCNN && ~213.18 & 0.7646~ & {\ul 0.8640} & 18.9990$^\dagger$ & {\ul 0.9652}~ & \textbf{0.6897} & 10.3337 & \textbf{0.4014} & 0.9075 & 0.9296 & 0.9683 & 0.3263\\
SAMT-MEF && 1552.58 & {\ul 0.8024}$^\dagger$ & 0.8596 & {\ul 19.7875}$^\dagger$ & 0.9076~ & 0.6389 & {\ul 8.9166} & 0.2872 & 0.9026 & 0.9343 & 0.9698 & 0.3412\\
\textbf{MobileMEF} && ~~~~0.88 & 0.7977~ & \textbf{0.8799} & \textbf{20.7976}~ & 0.8957~ & 0.6325 & \textbf{8.8717} & {\ul 0.3306} & \textbf{0.9320} & \textbf{0.9622} & \textbf{0.9771} & \textbf{0.3036} \\ \hline

\multicolumn{6}{l}{\footnotesize $^\dagger$The metric is also used in the method loss function formulation.}
\end{tabular}
}
\label{tab:comp_results}
\end{table*}

\def\imgA{101}
\def\imgB{10}
\def\imgC{31}
\def\sizeH{0.15\textwidth}
\def\sizeW{0.24\textwidth}

\begin{figure*}[!t]
\centering
\subfloat[][Inputs]{\includegraphics[width=\sizeW, height=\sizeH]{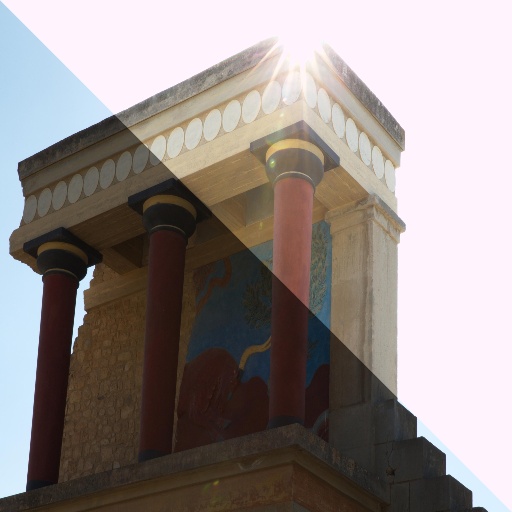}}
~
\subfloat[][HoLoCo]{\includegraphics[width=\sizeW, height=\sizeH]{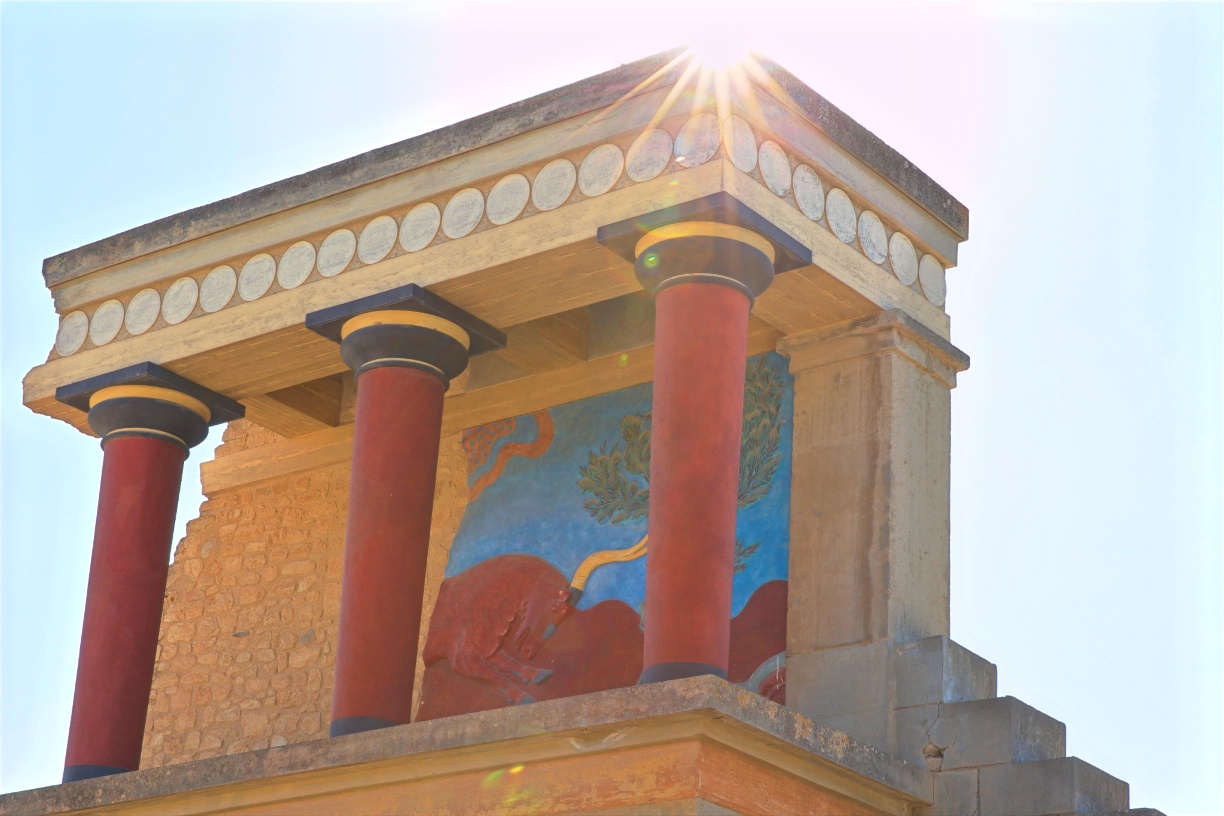}}
~
\subfloat[][MEFLUT]{\includegraphics[width=\sizeW, height=\sizeH]{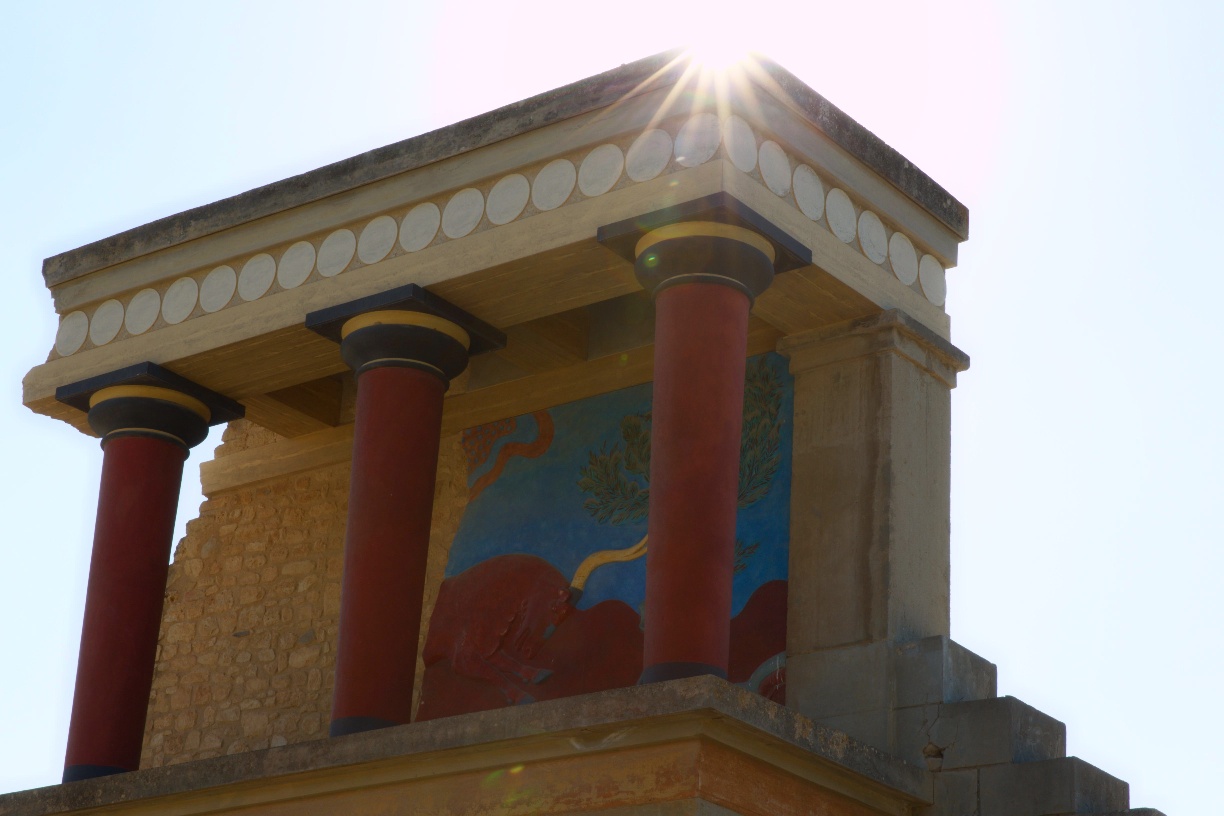}}
~
\subfloat[][\textbf{MobileMEF}]{\includegraphics[width=\sizeW, height=\sizeH]{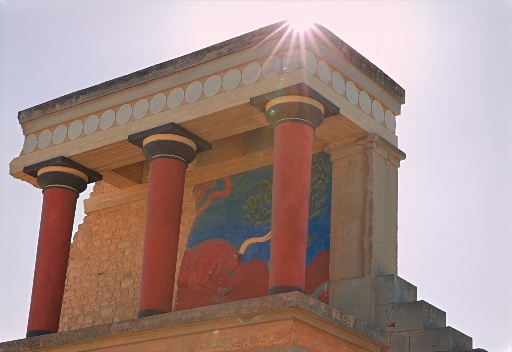}}
\\
\subfloat[][SAMT-MEF]{\includegraphics[width=\sizeW, height=\sizeH]{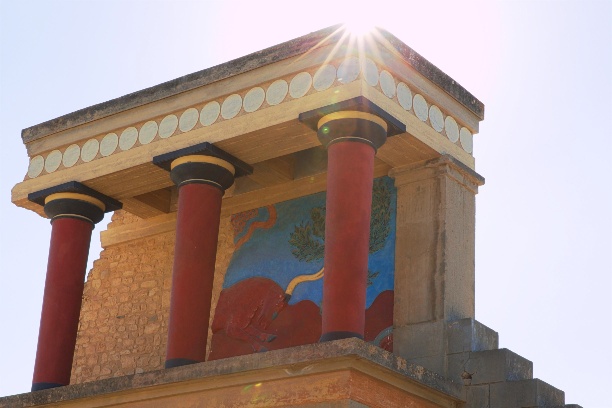}}
~
\subfloat[][IFCNN]{\includegraphics[width=\sizeW, height=\sizeH]{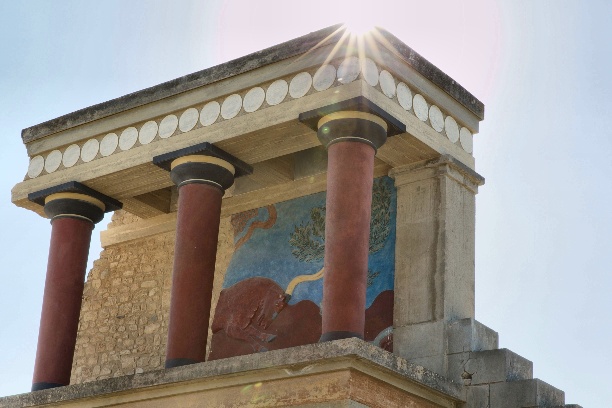}}
~
\subfloat[][TransMEF]{\includegraphics[width=\sizeW, height=\sizeH]{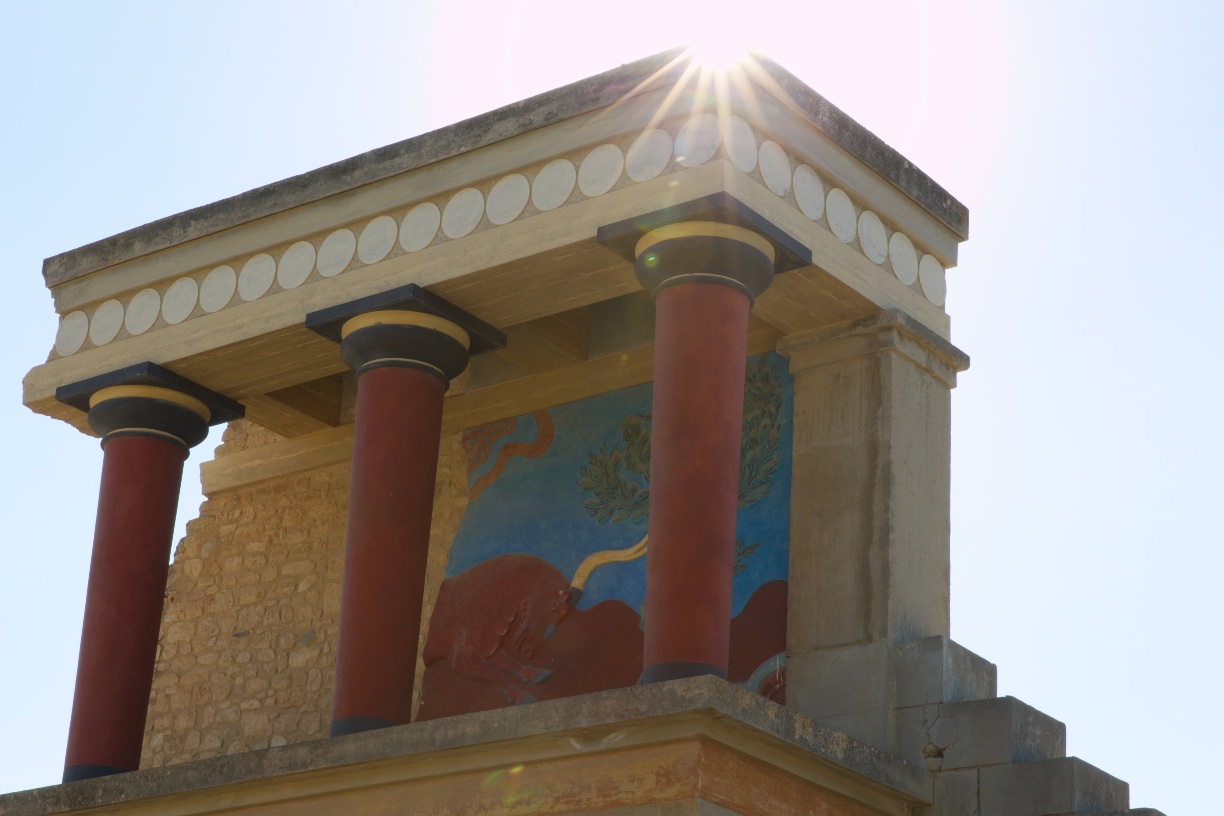}}
~
\subfloat[][Ground-truth]{\includegraphics[width=\sizeW, height=\sizeH]{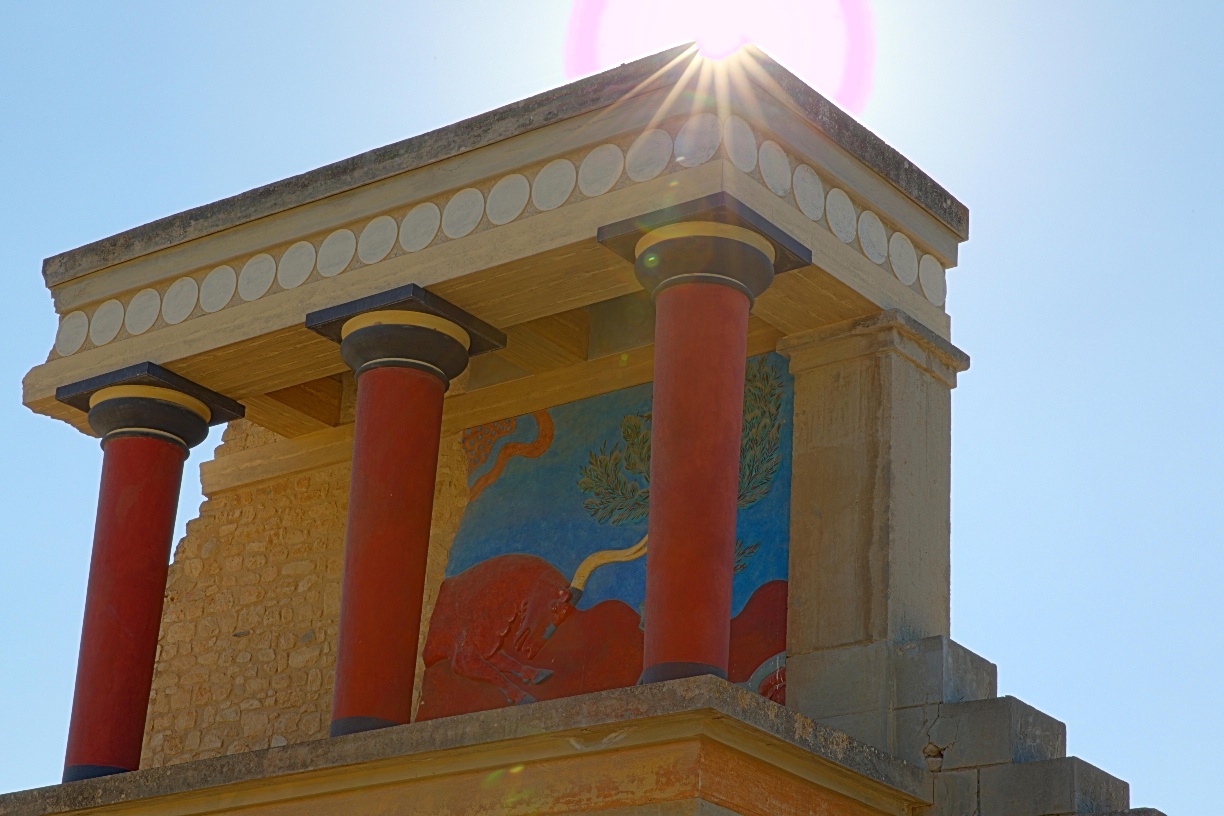}}
\\
\subfloat[][Inputs]{\includegraphics[width=\sizeW, height=\sizeH]{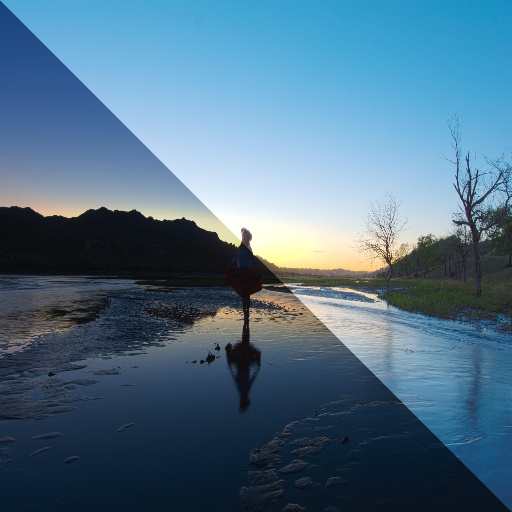}}
~
\subfloat[][HoLoCo]{\includegraphics[width=\sizeW, height=\sizeH]{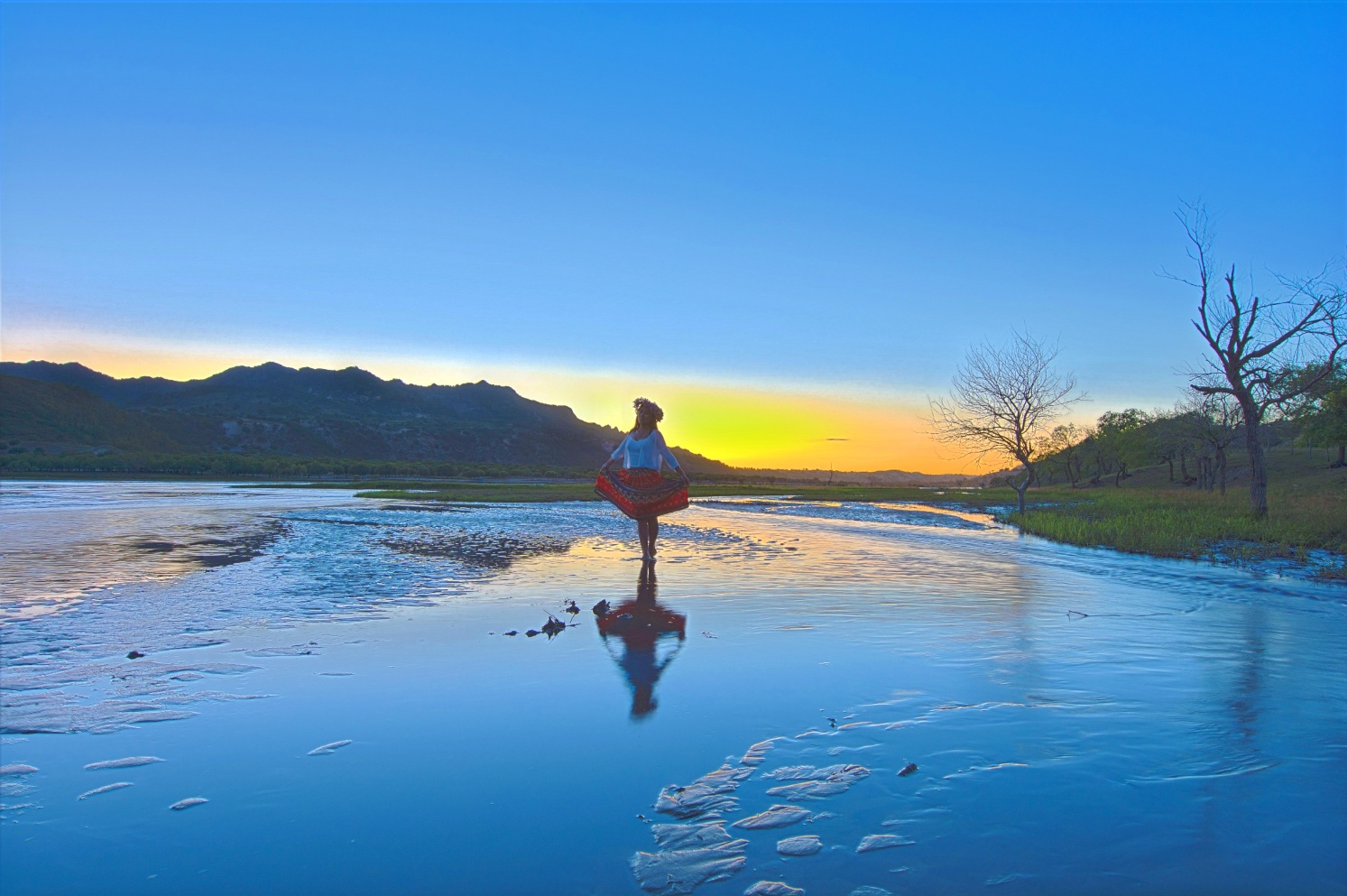}}
~
\subfloat[][MEFLUT]{\includegraphics[width=\sizeW, height=\sizeH]{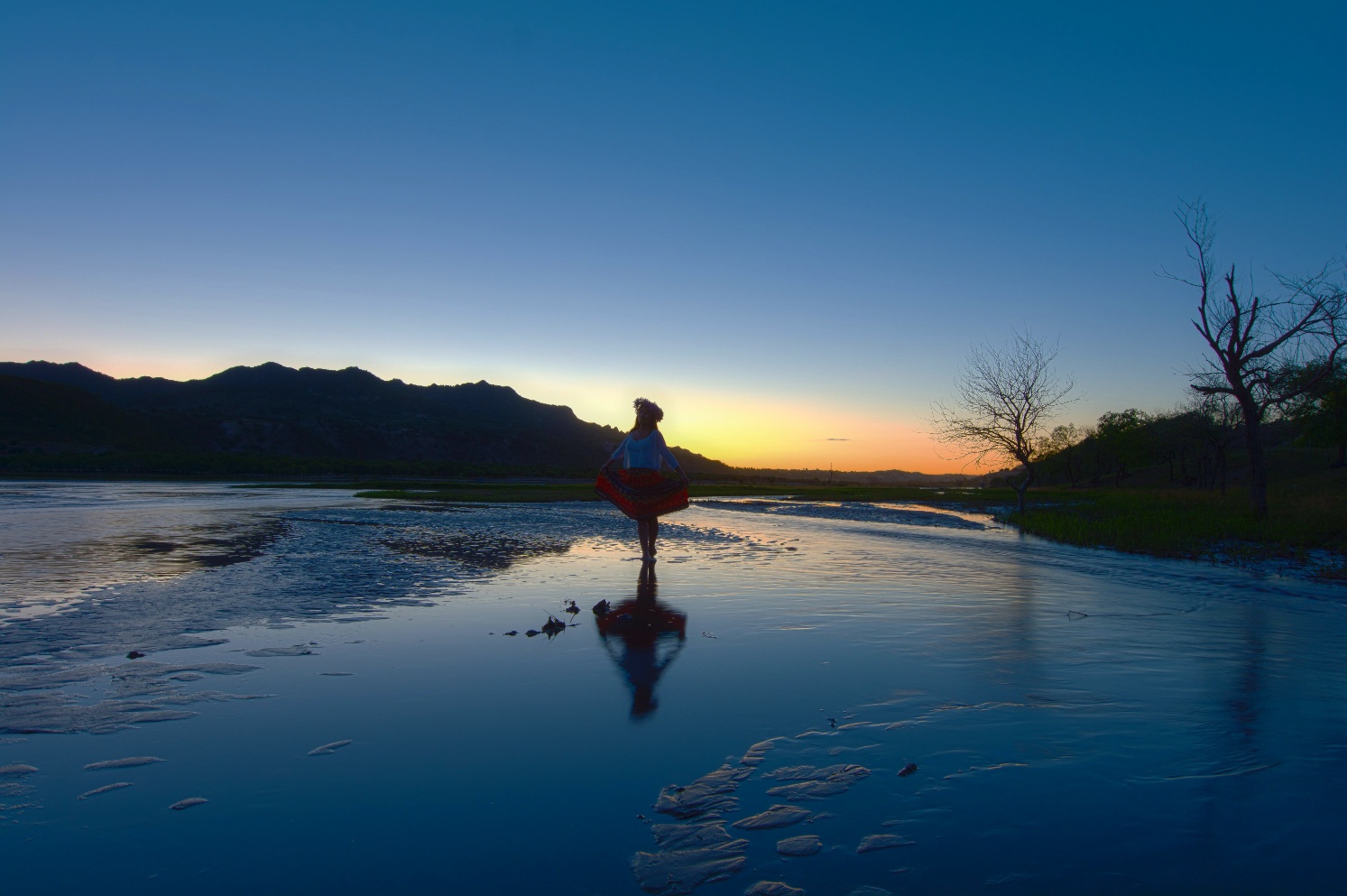}}
~
\subfloat[][\textbf{MobileMEF}]{\includegraphics[width=\sizeW, height=\sizeH]{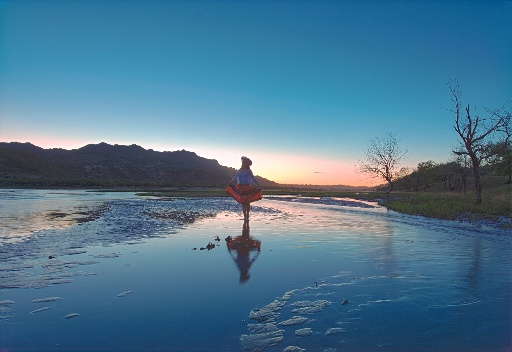}}
\\
\subfloat[][SAMT-MEF]{\includegraphics[width=\sizeW, height=\sizeH]{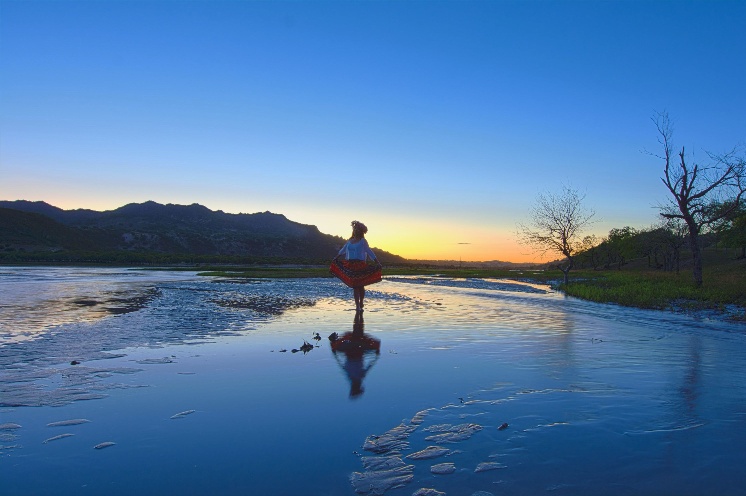}}
~
\subfloat[][IFCNN]{\includegraphics[width=\sizeW, height=\sizeH]{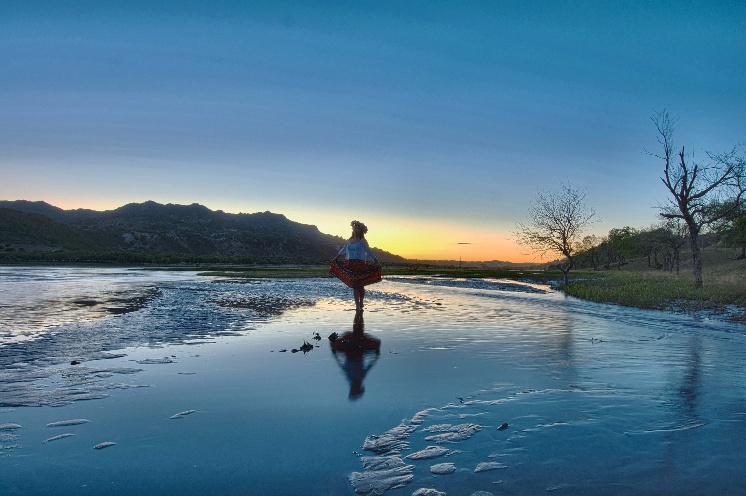}}
~
\subfloat[][TransMEF]{\includegraphics[width=\sizeW, height=\sizeH]{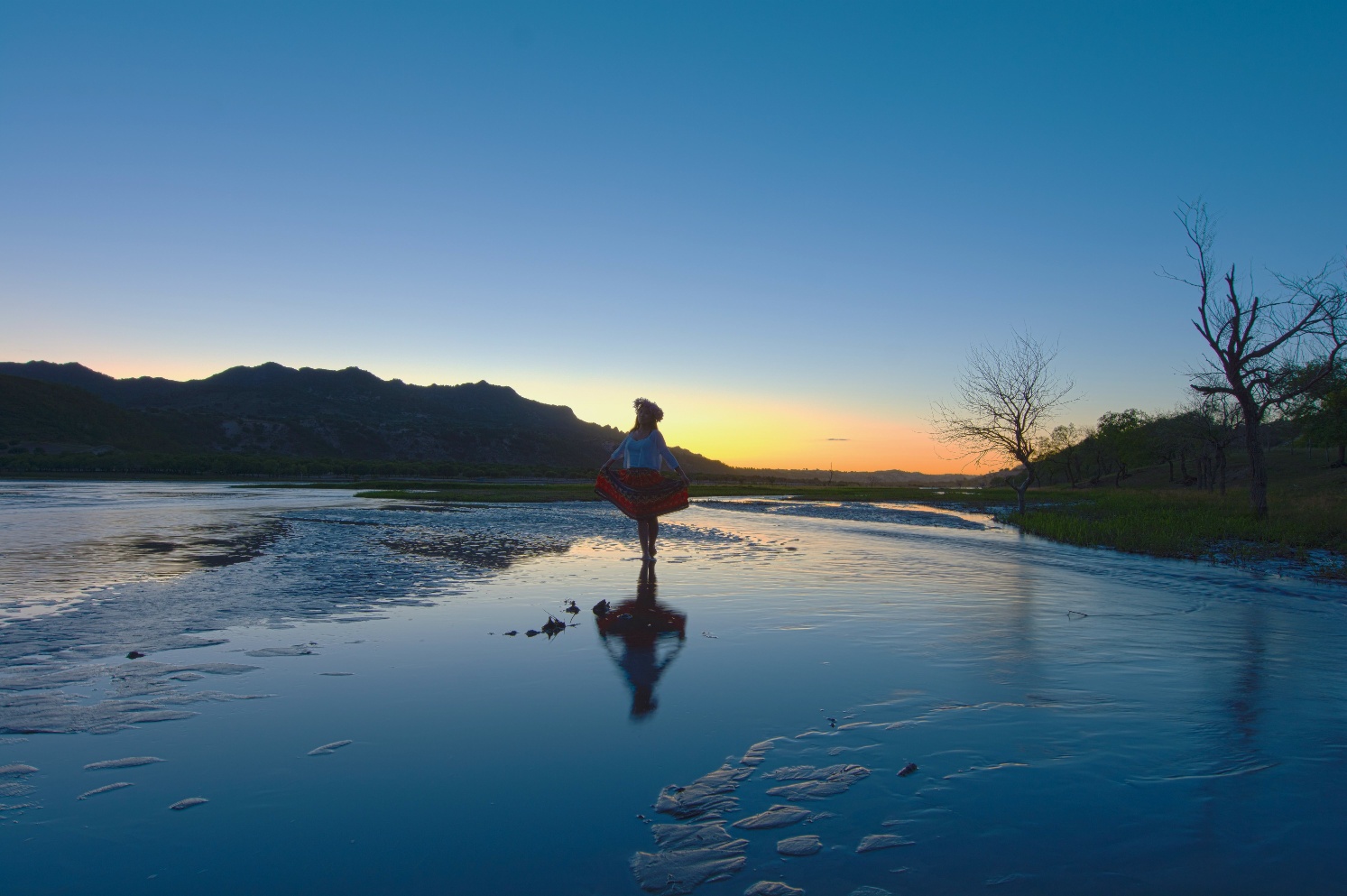}}
~
\subfloat[][Ground-truth]{\includegraphics[width=\sizeW, height=\sizeH]{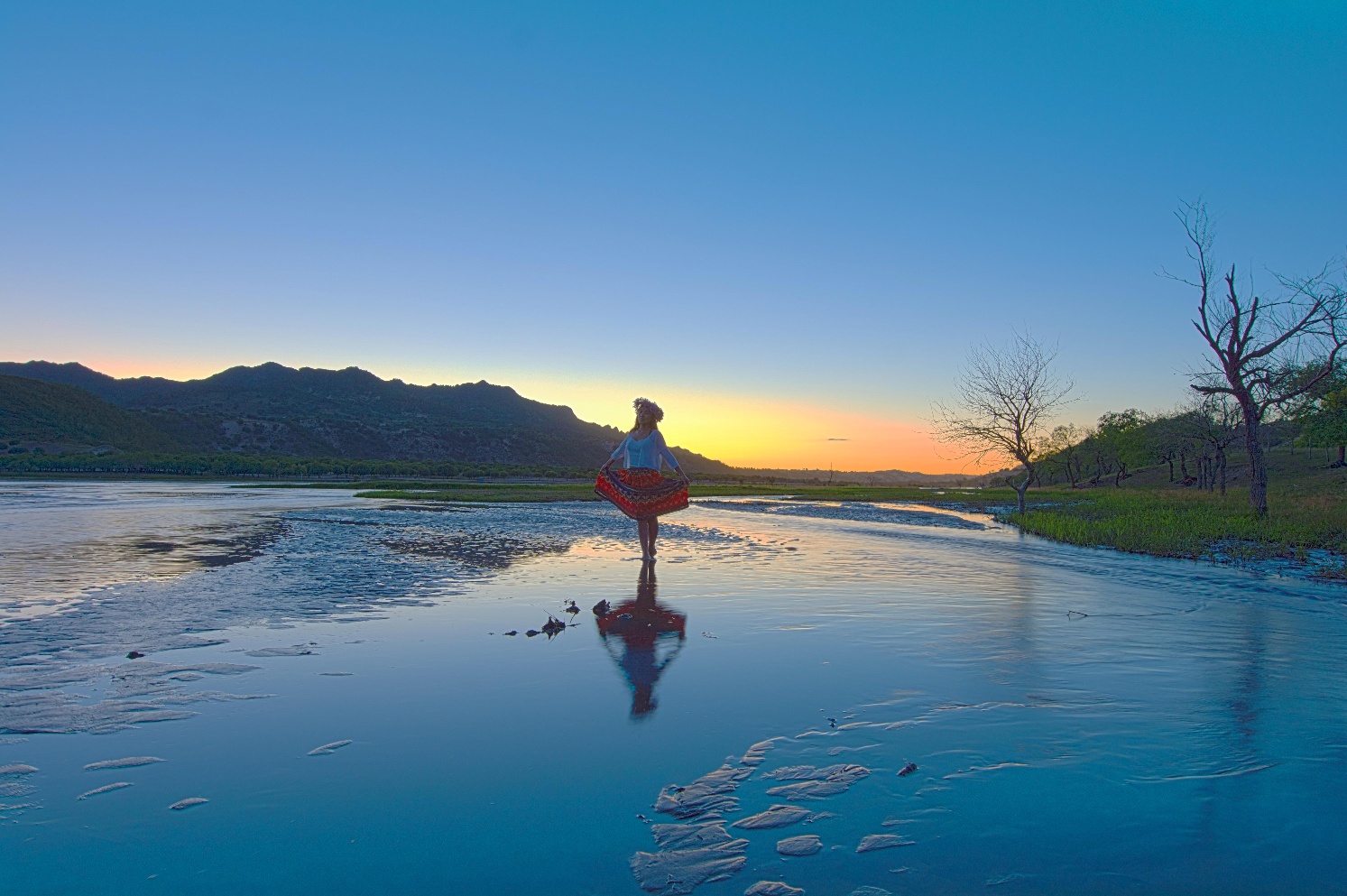}}
\\
\subfloat[][Inputs]{\includegraphics[width=\sizeW, height=\sizeH]{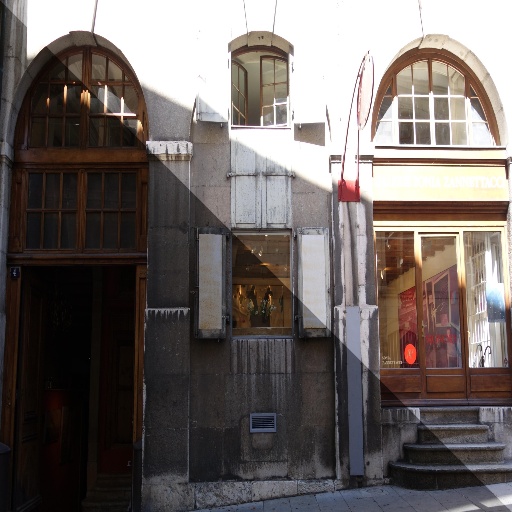}}
~
\subfloat[][HoLoCo]{\includegraphics[width=\sizeW, height=\sizeH]{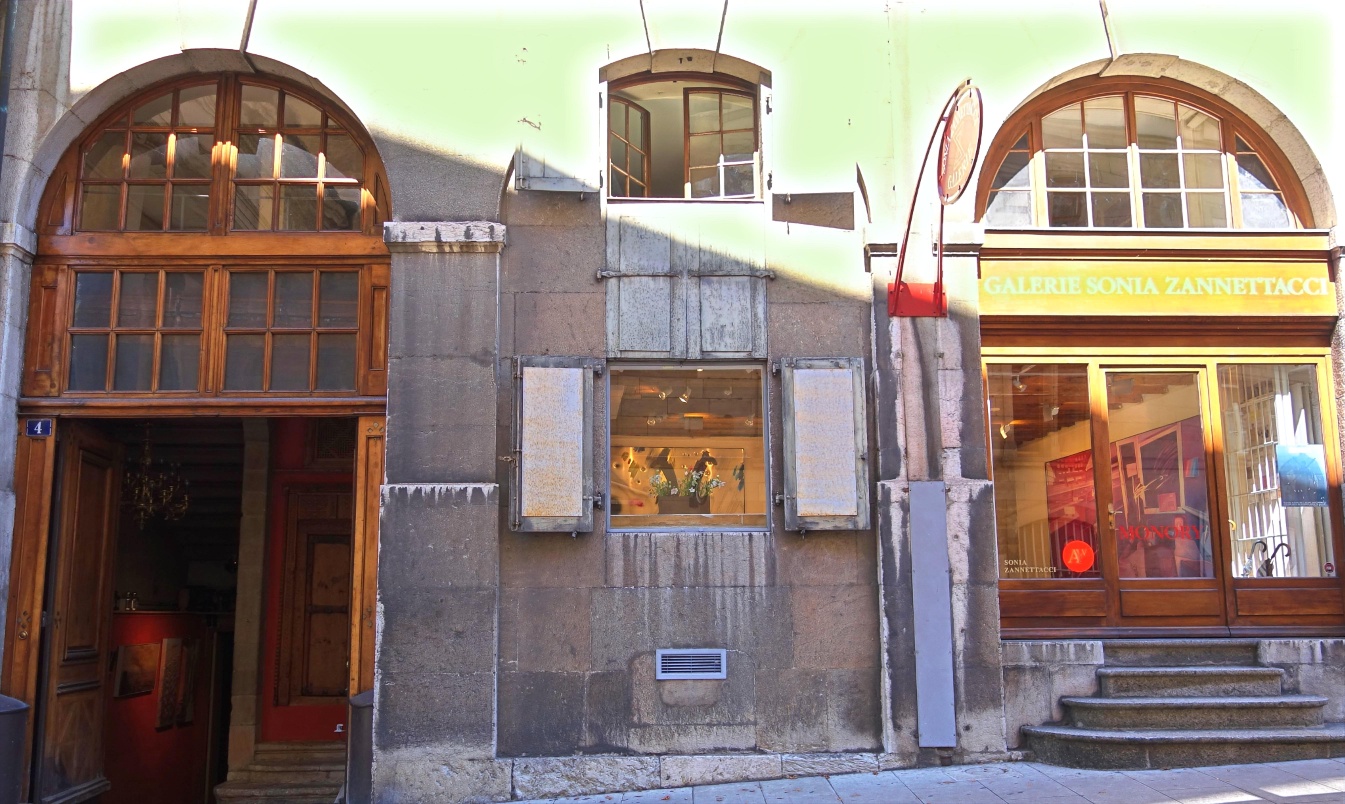}}
~
\subfloat[][MEFLUT]{\includegraphics[width=\sizeW, height=\sizeH]{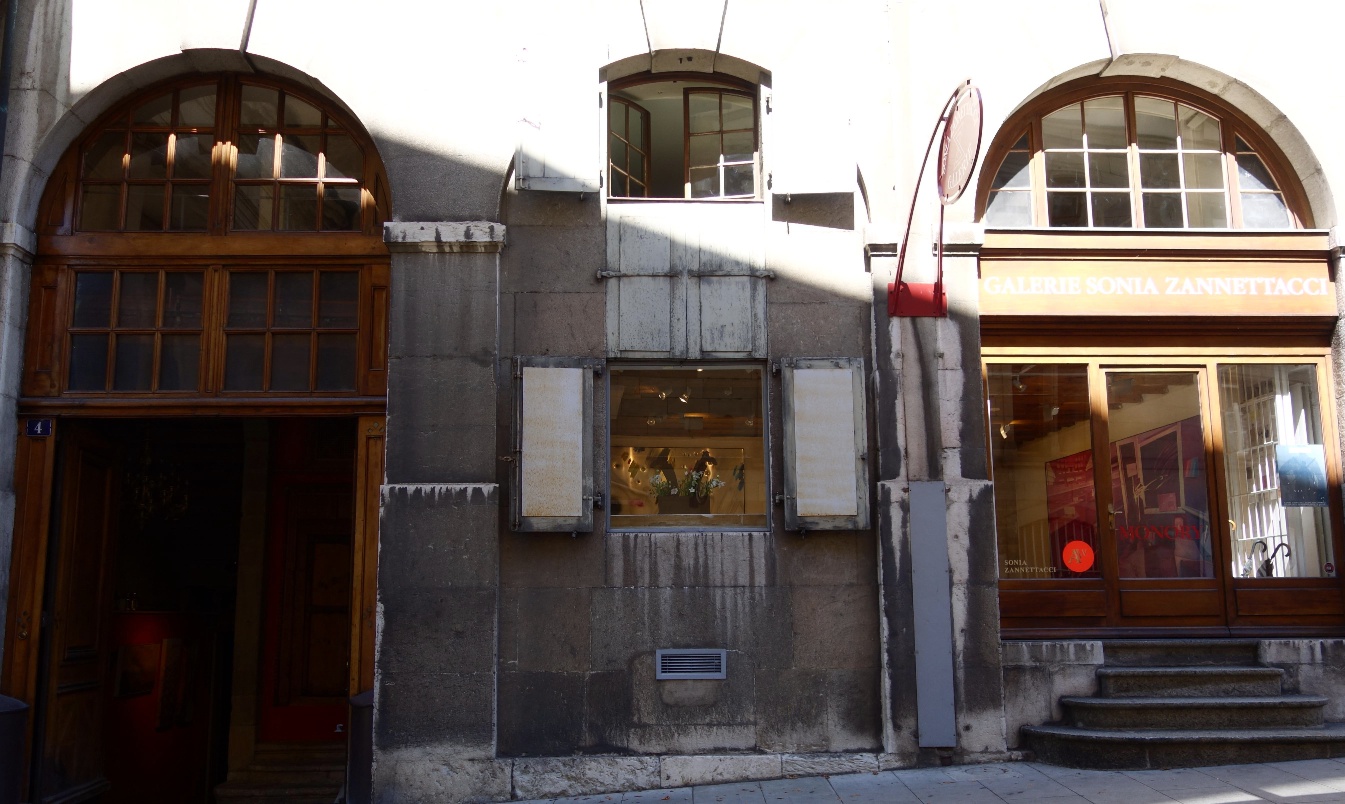}}
~
\subfloat[][\textbf{MobileMEF}]{\includegraphics[width=\sizeW, height=\sizeH]{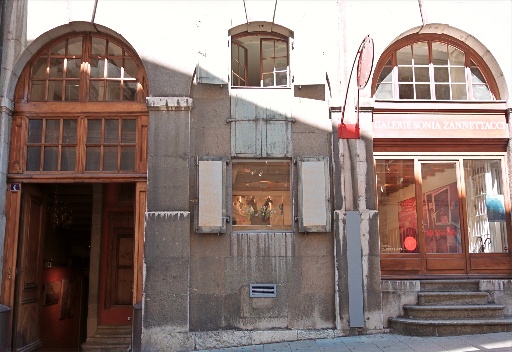}}
\\
\subfloat[][SAMT-MEF]{\includegraphics[width=\sizeW, height=\sizeH]{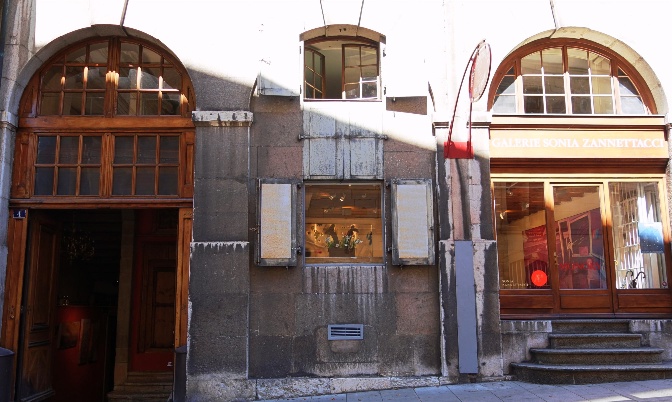}}
~
\subfloat[][IFCNN]{\includegraphics[width=\sizeW, height=\sizeH]{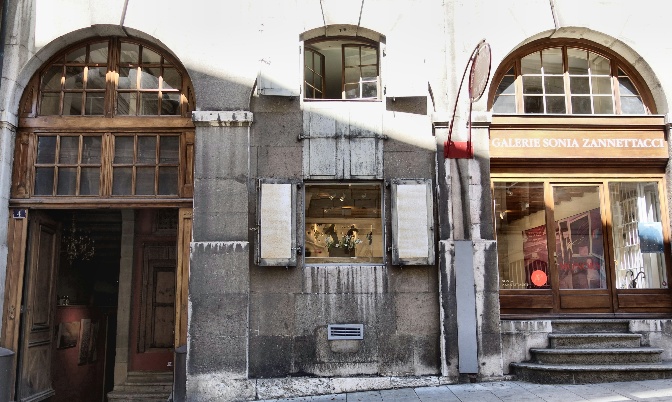}}
~
\subfloat[][TransMEF]{\includegraphics[width=\sizeW, height=\sizeH]{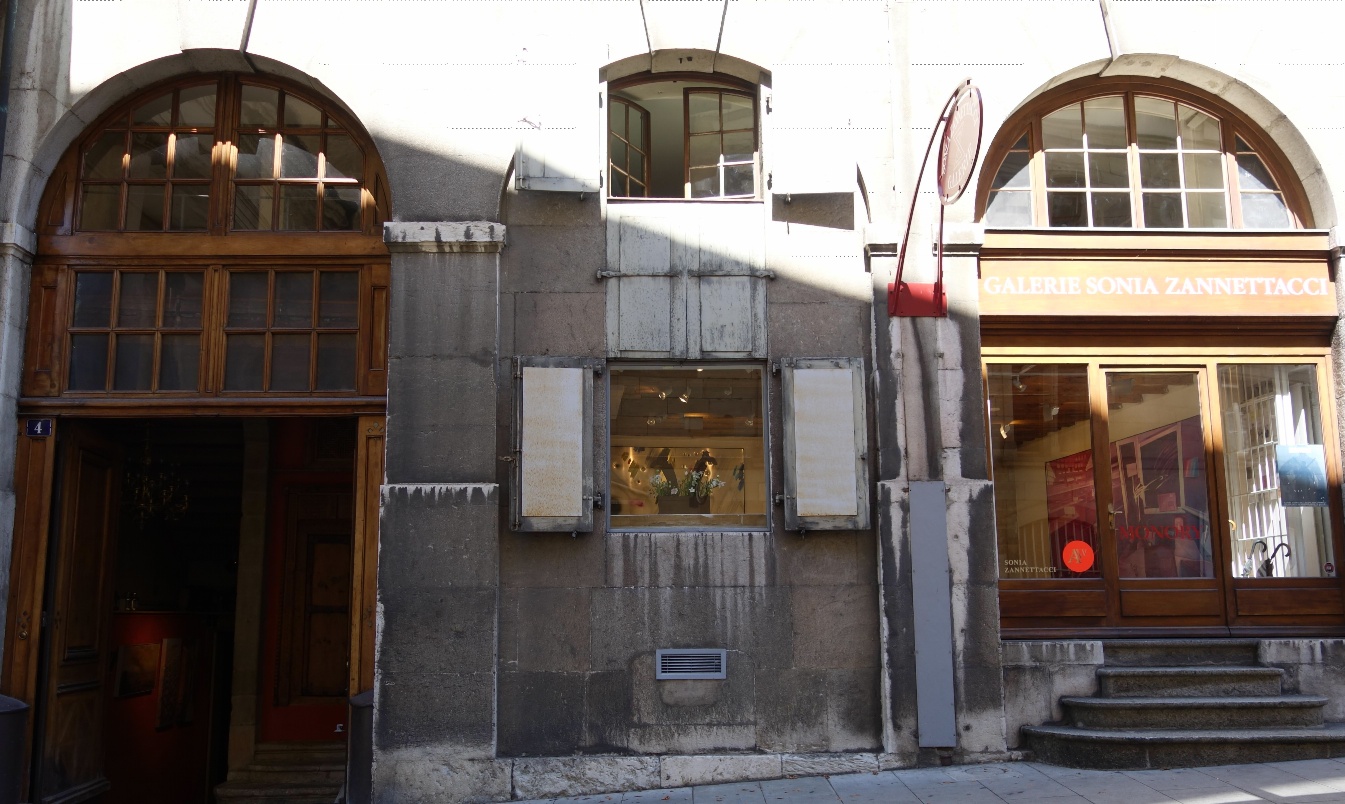}}
~
\subfloat[][Ground-truth]{\includegraphics[width=\sizeW, height=\sizeH]{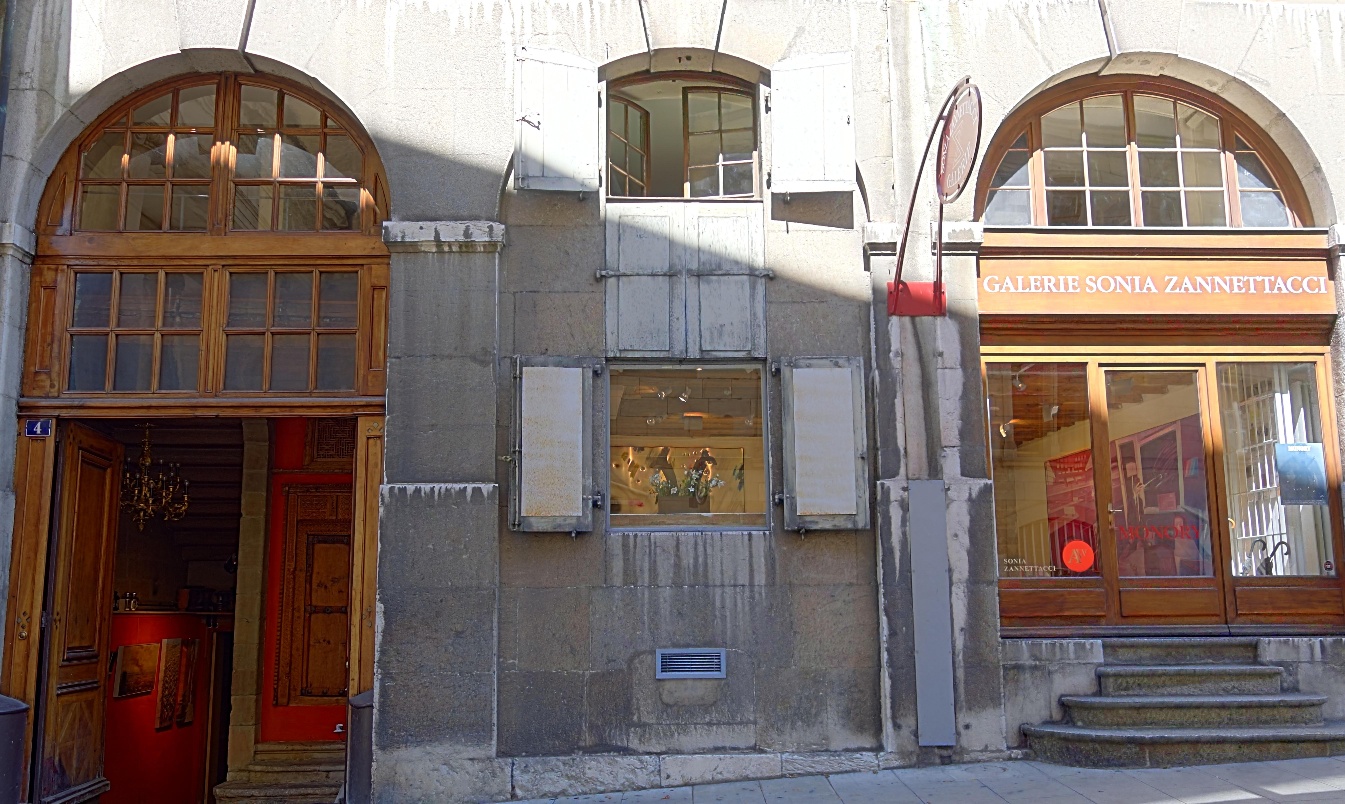}}
\caption{Visual comparison of MobileMEF with SOTA methods using EV +1 and -1 as input frames.}
\label{fig:qualitative}
\end{figure*}

\begin{figure*}[!t]
\centering
\subfloat[][Inputs]{\includegraphics[width=\sizeW, height=\sizeH]{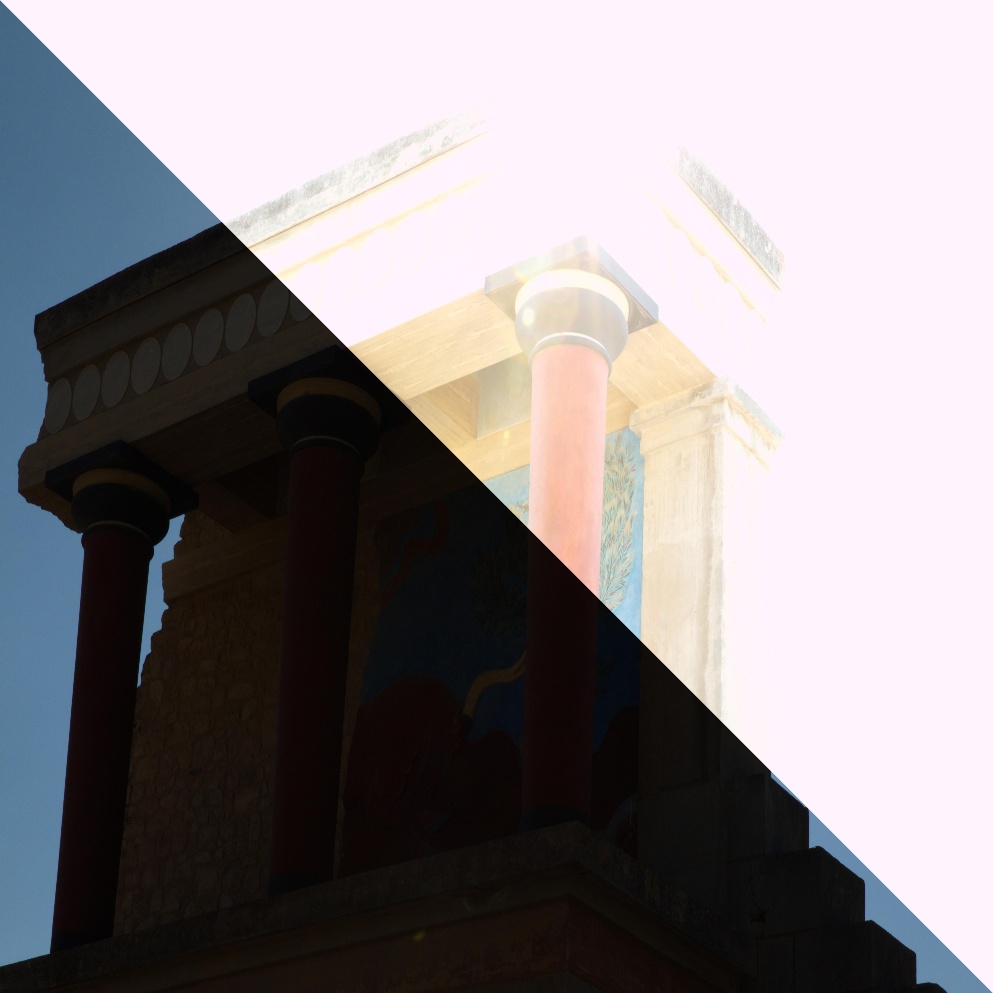}}
~
\subfloat[][HoLoCo]{\includegraphics[width=\sizeW, height=\sizeH]{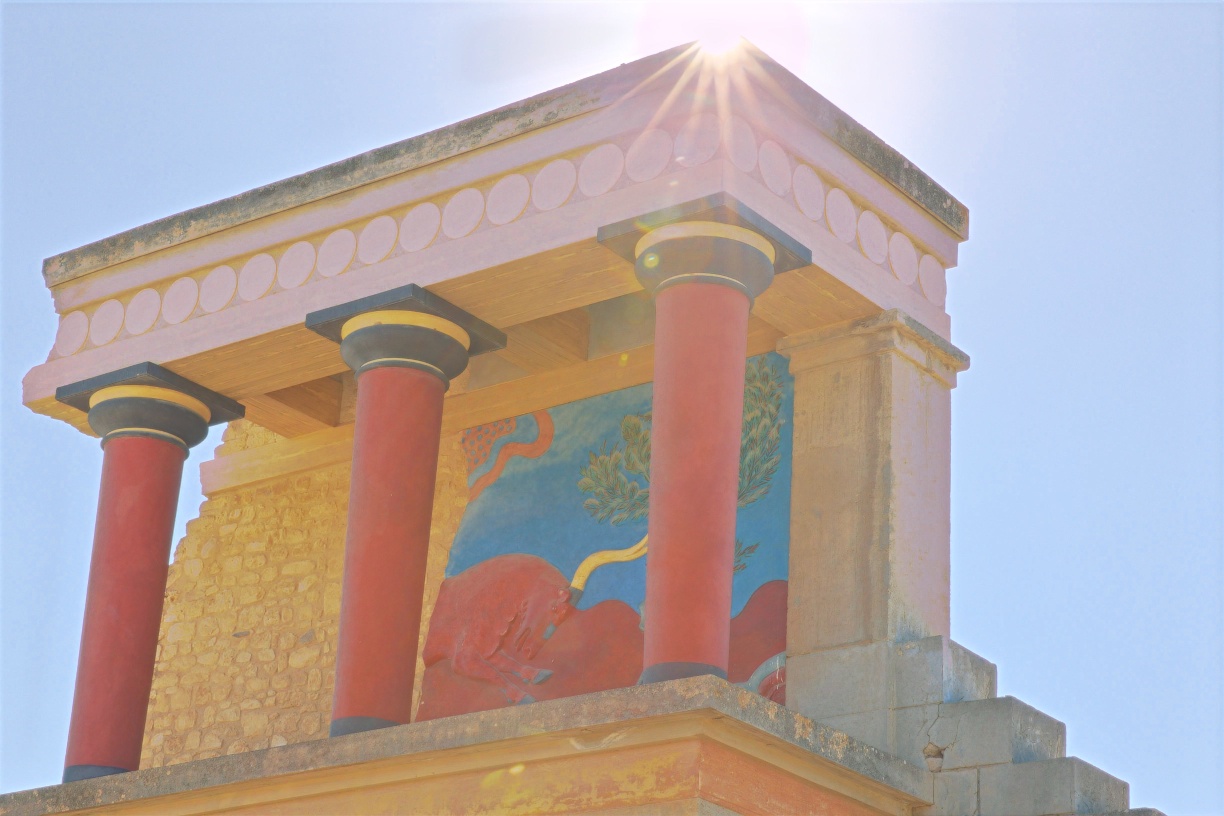}}
~
\subfloat[][MEFLUT]{\includegraphics[width=\sizeW, height=\sizeH]{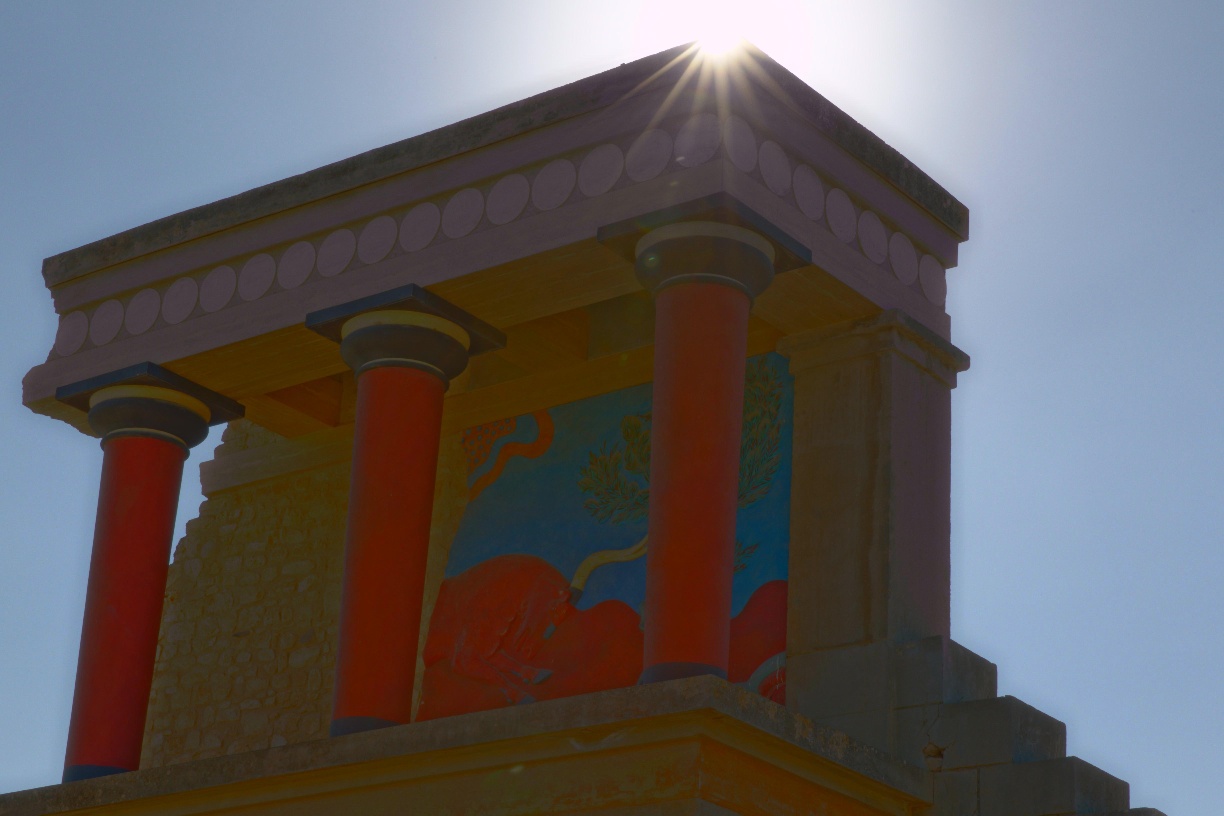}}
~
\subfloat[][\textbf{MobileMEF}]{\includegraphics[width=\sizeW, height=\sizeH]{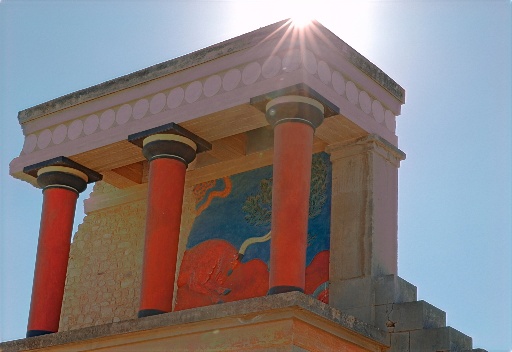}}
\\
\subfloat[][SAMT-MEF]{\includegraphics[width=\sizeW, height=\sizeH]{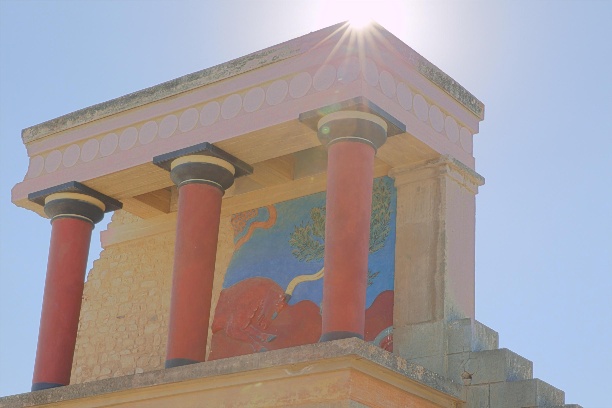}}
~
\subfloat[][IFCNN]{\includegraphics[width=\sizeW, height=\sizeH]{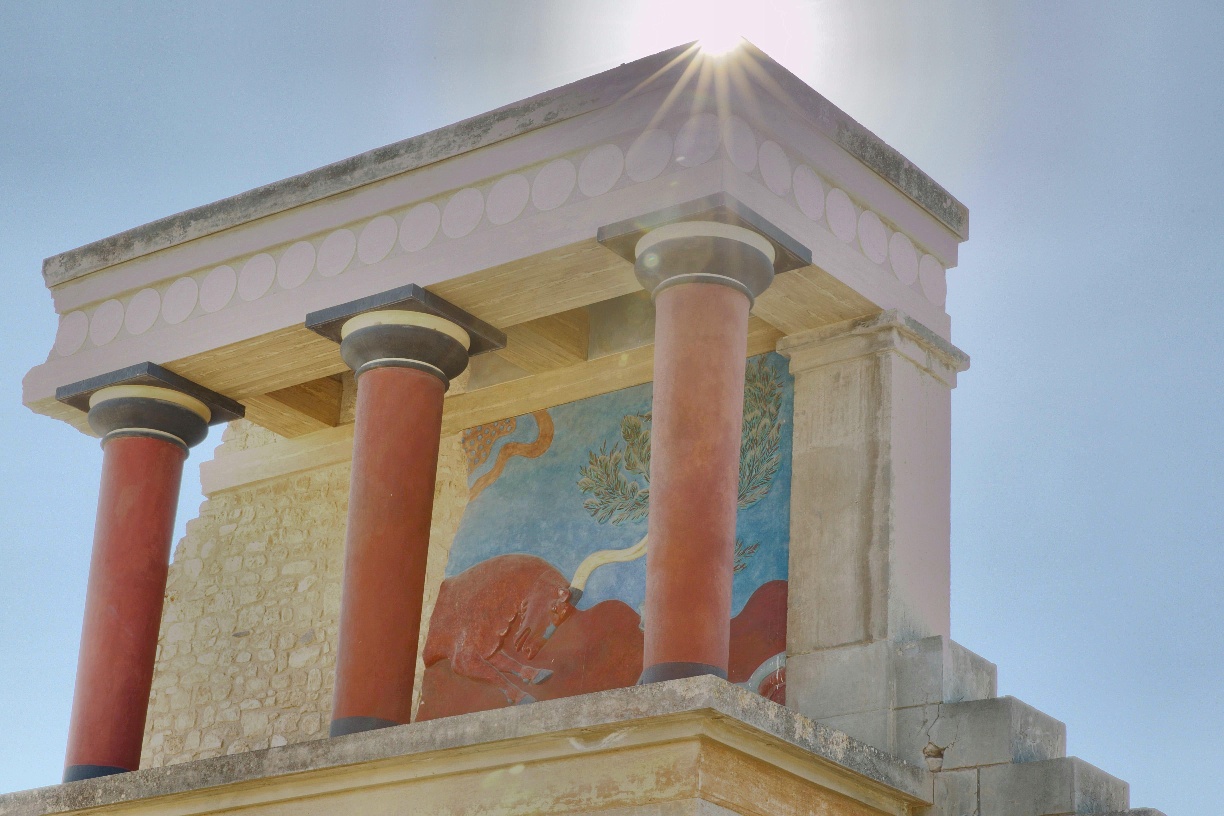}}
~
\subfloat[][TransMEF]{\includegraphics[width=\sizeW, height=\sizeH]{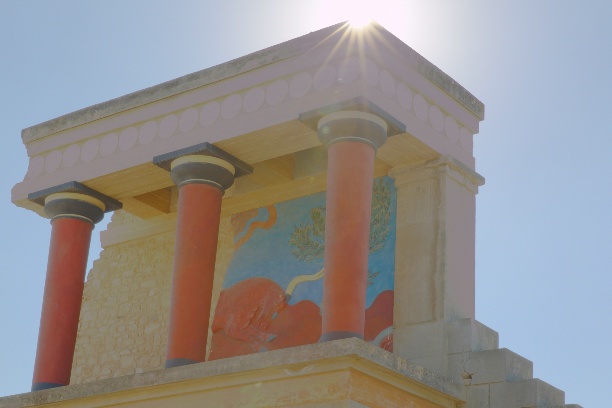}}
~
\subfloat[][Ground-truth]{\includegraphics[width=\sizeW, height=\sizeH]{images/GT/\imgA.jpg}}
\\
\subfloat[][Inputs]{\includegraphics[width=\sizeW, height=\sizeH]{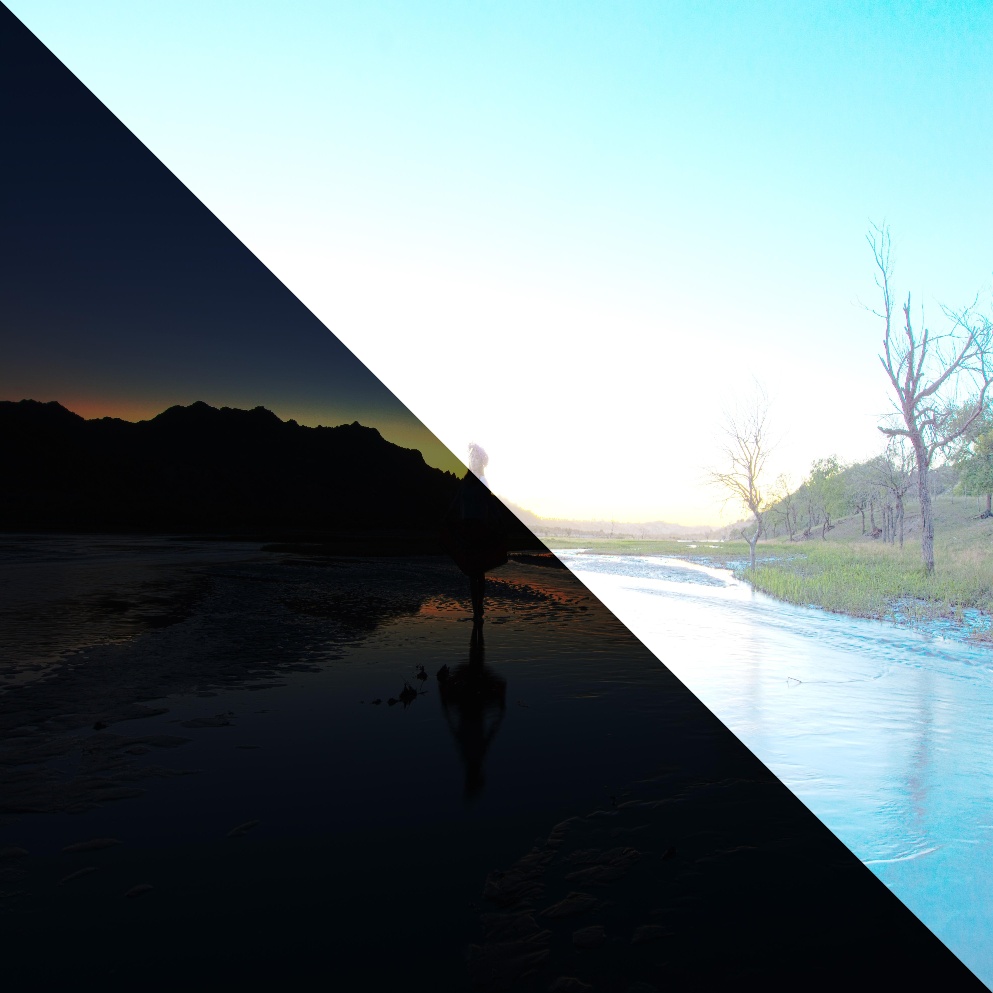}}
~
\subfloat[][HoLoCo]{\includegraphics[width=\sizeW, height=\sizeH]{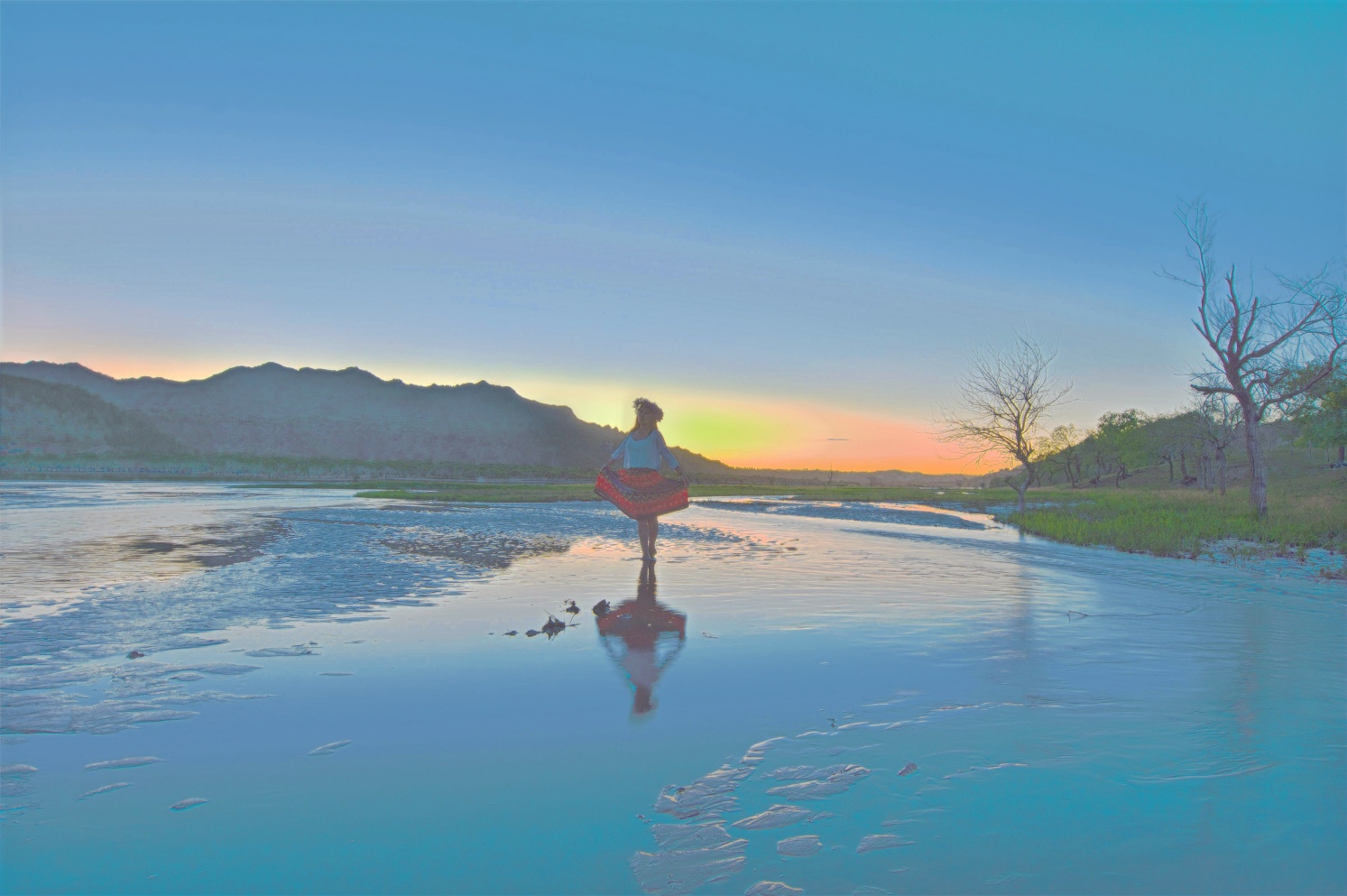}}
~
\subfloat[][MEFLUT]{\includegraphics[width=\sizeW, height=\sizeH]{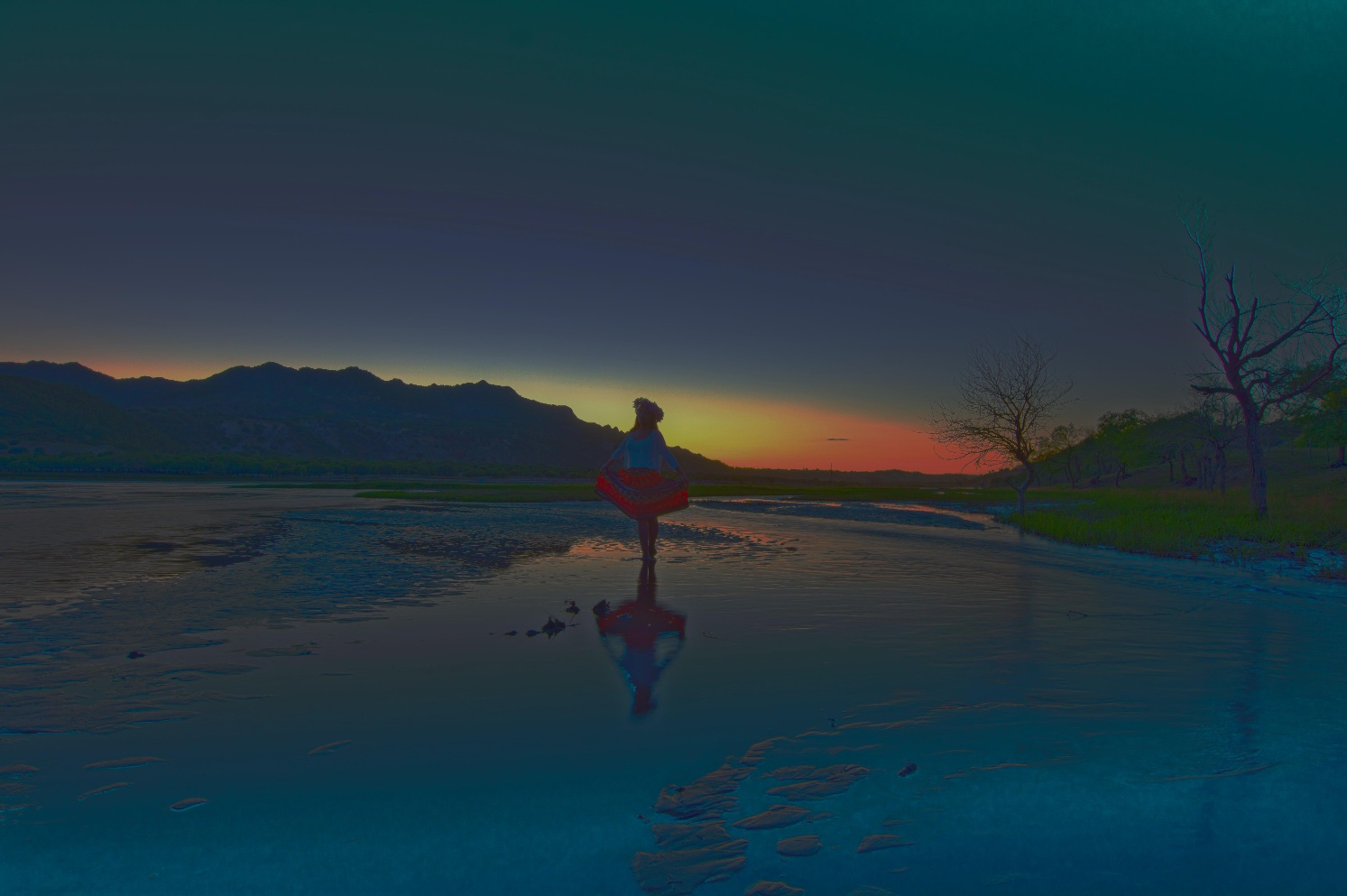}}
~
\subfloat[][\textbf{MobileMEF}]{\includegraphics[width=\sizeW, height=\sizeH]{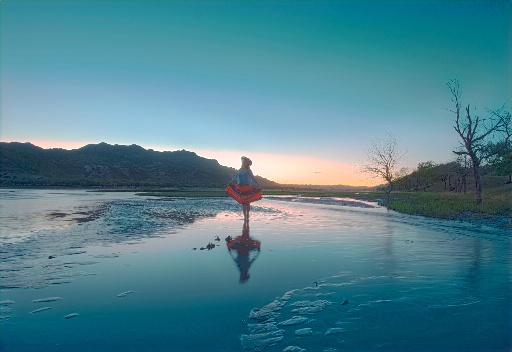}}
\\
\subfloat[][SAMT-MEF]{\includegraphics[width=\sizeW, height=\sizeH]{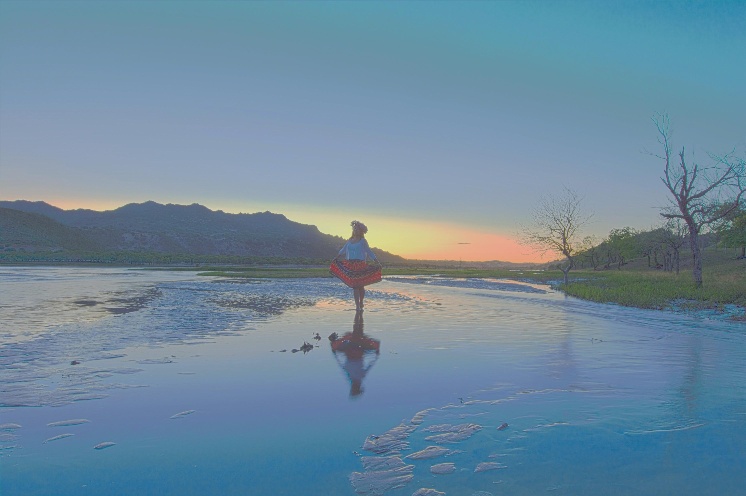}}
~
\subfloat[][IFCNN]{\includegraphics[width=\sizeW, height=\sizeH]{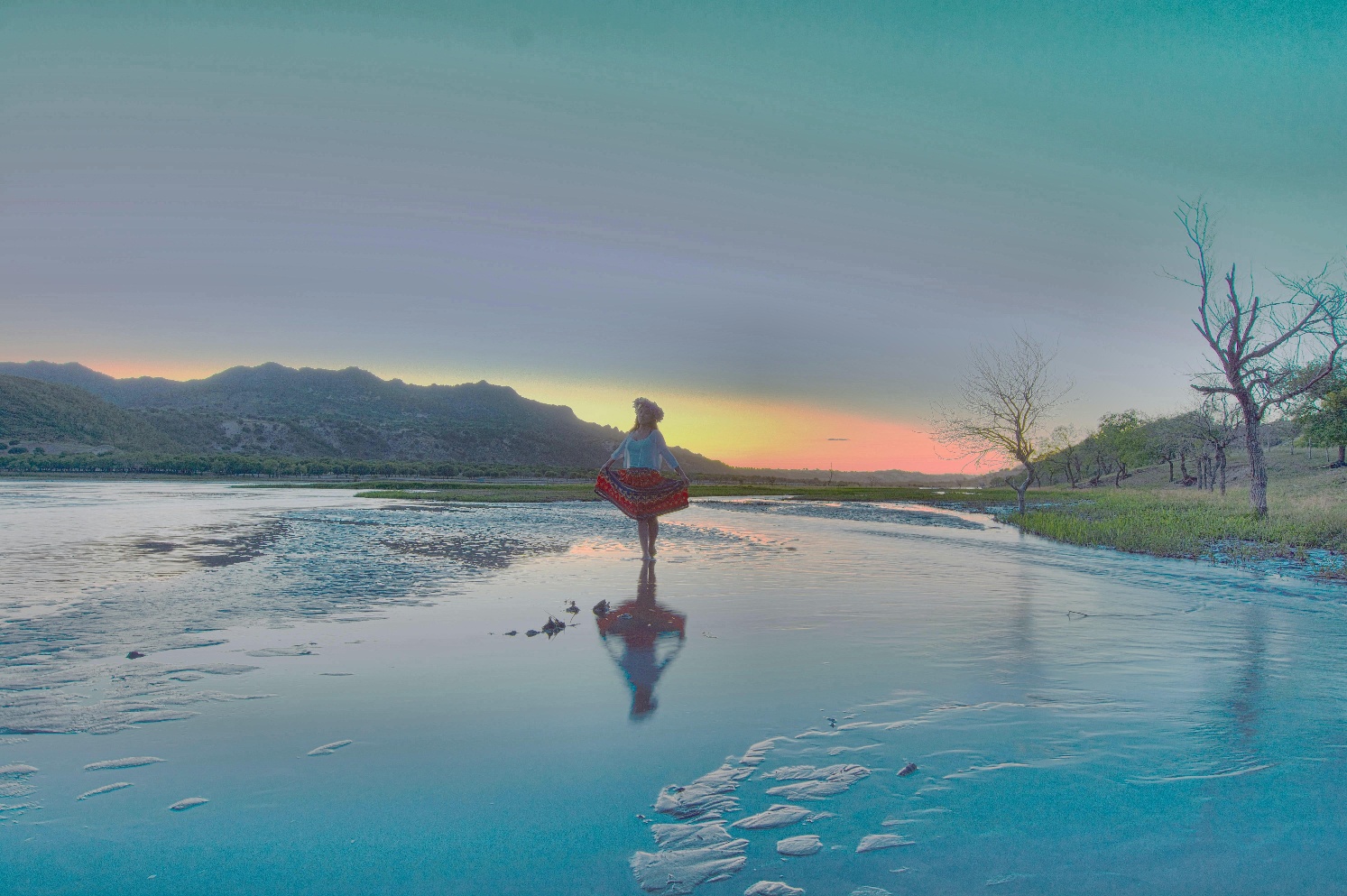}}
~
\subfloat[][TransMEF]{\includegraphics[width=\sizeW, height=\sizeH]{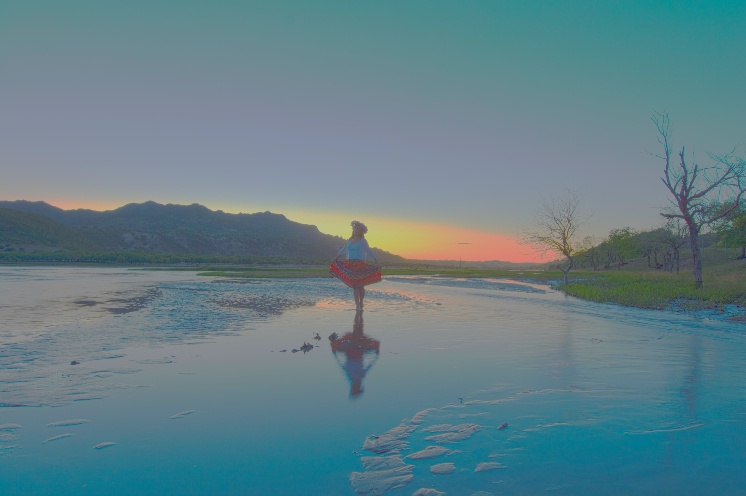}}
~
\subfloat[][Ground-truth]{\includegraphics[width=\sizeW, height=\sizeH]{images/GT/\imgB.jpg}}
\\
\subfloat[][Inputs]{\includegraphics[width=\sizeW, height=\sizeH]{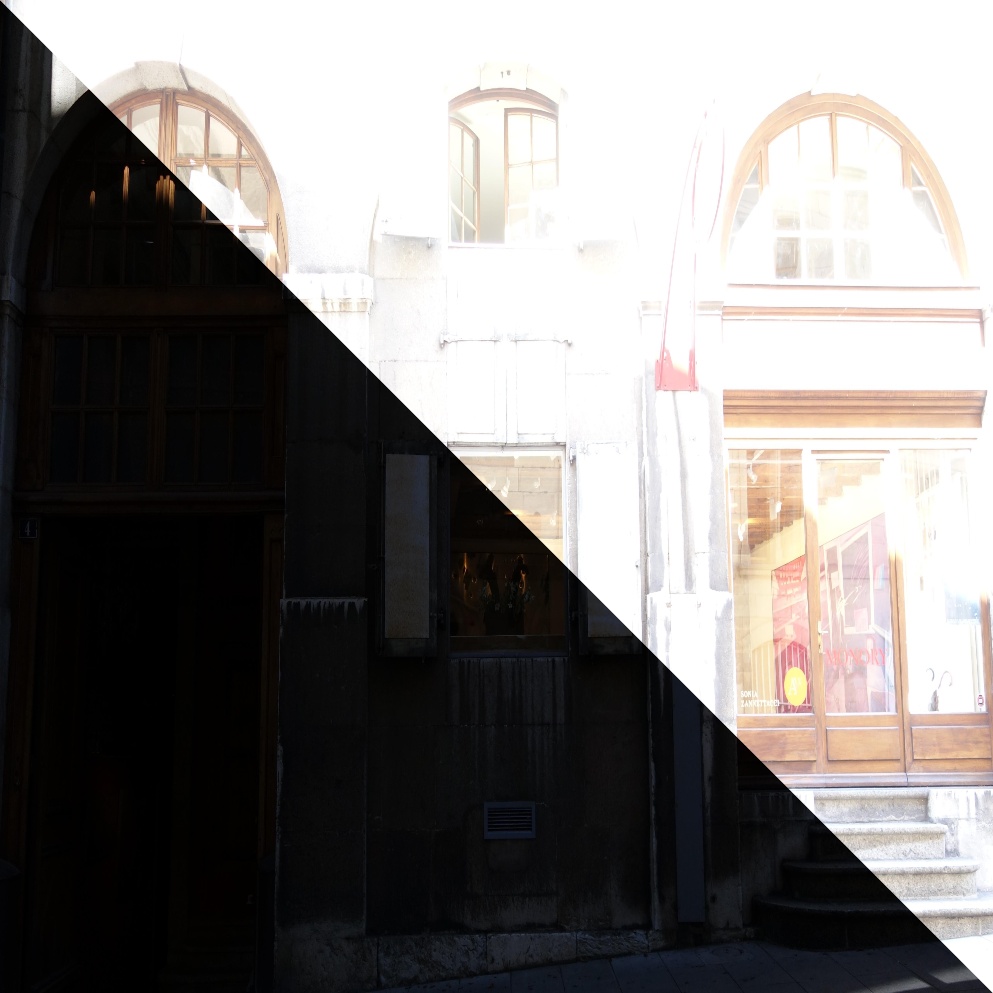}}
~
\subfloat[][HoLoCo]{\includegraphics[width=\sizeW, height=\sizeH]{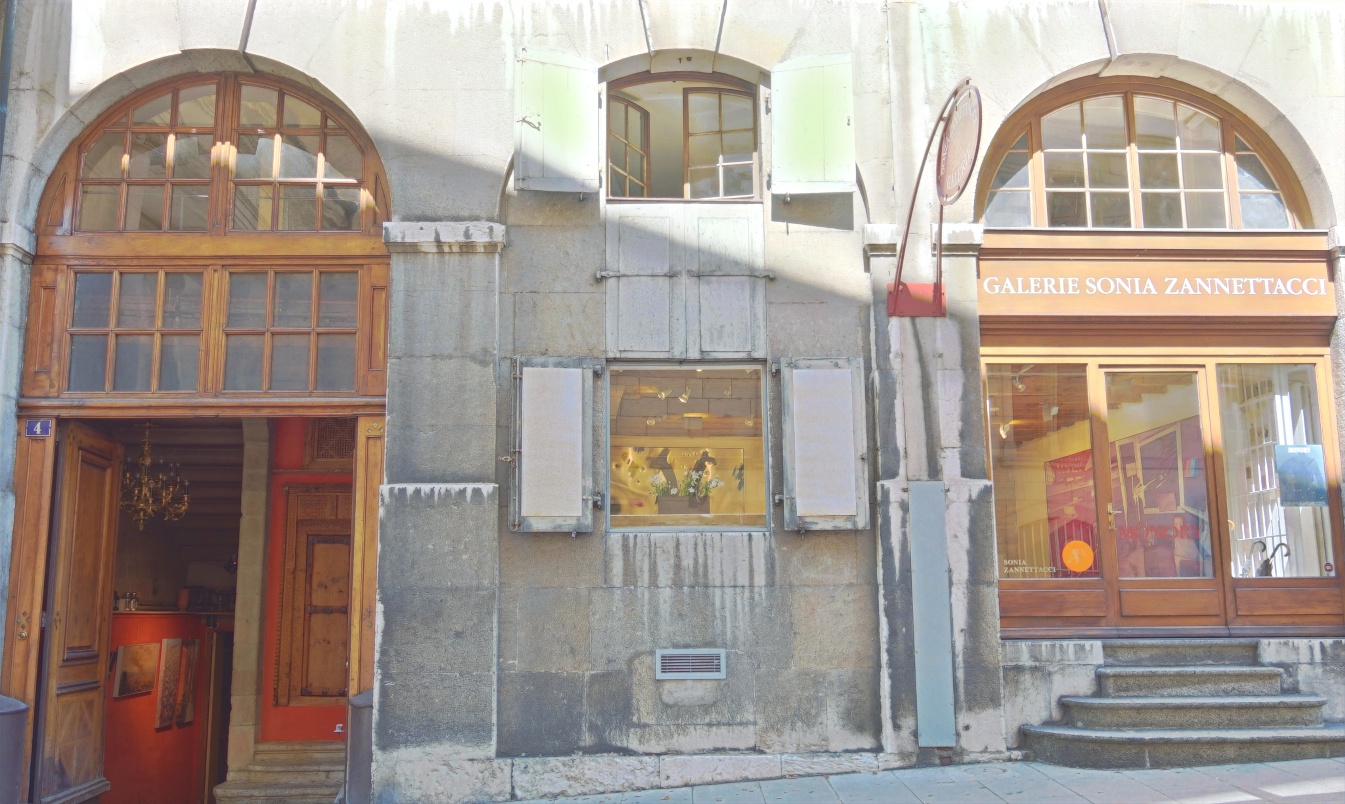}}
~
\subfloat[][MEFLUT]{\includegraphics[width=\sizeW, height=\sizeH]{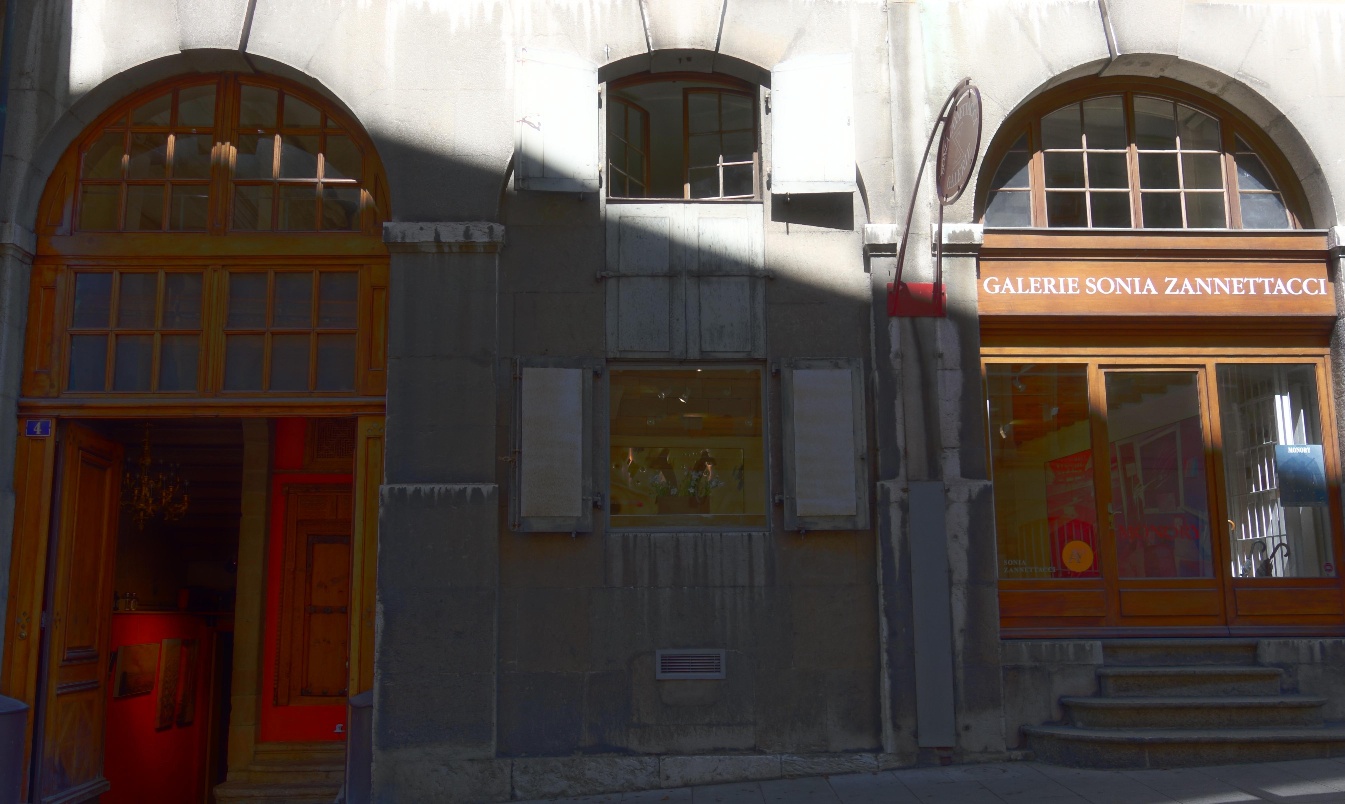}}
~
\subfloat[][\textbf{MobileMEF}]{\includegraphics[width=\sizeW, height=\sizeH]{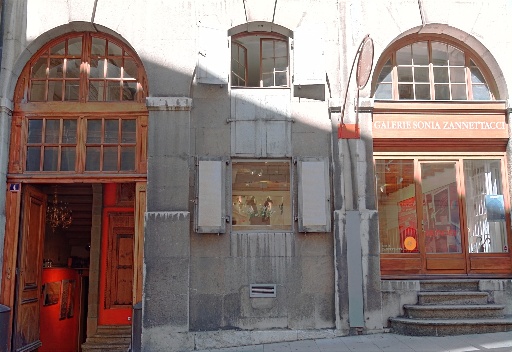}}
\\
\subfloat[][SAMT-MEF]{\includegraphics[width=\sizeW, height=\sizeH]{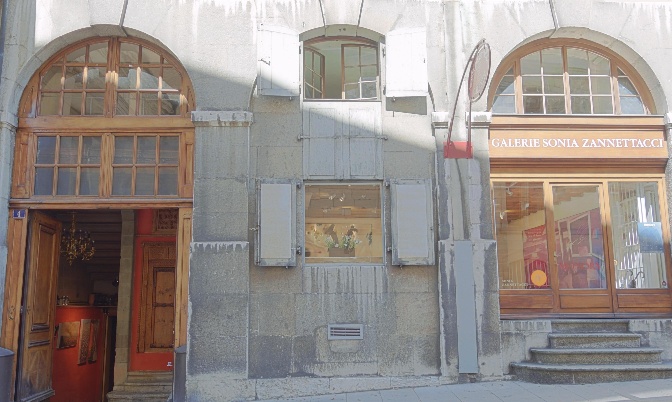}}
~
\subfloat[][IFCNN]{\includegraphics[width=\sizeW, height=\sizeH]{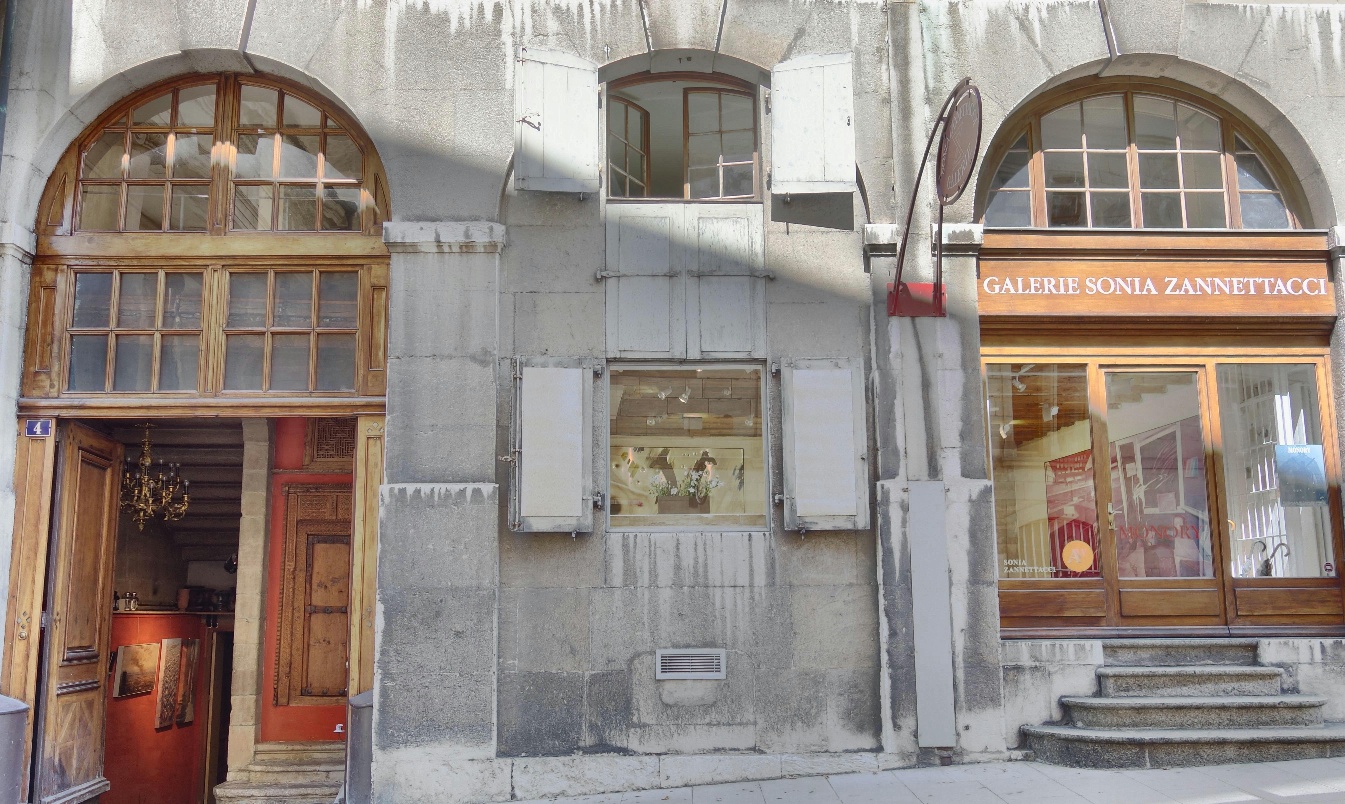}}
~
\subfloat[][TransMEF]{\includegraphics[width=\sizeW, height=\sizeH]{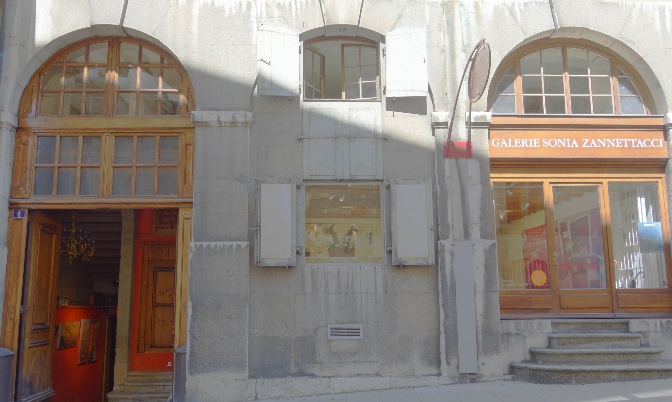}}
~
\subfloat[][Ground-truth]{\includegraphics[width=\sizeW, height=\sizeH]{images/GT/\imgC.jpg}}
\caption{Visual comparison of MobileMEF with SOTA methods using the most under and over exposed input frames.}
\label{fig:qualitative2}
\end{figure*}

Regarding our visual results presented in Figs.~\ref{fig:qualitative}~and~\ref{fig:qualitative2}, our method demonstrates noticeable improvements in visual quality. It consistently delivers images with better contrast and detail preservation, closely resembling the ground truth, especially when compared to alternatives like HoLoCo~\cite{holoco}, MEFLUT~\cite{meflut}, and IFCNN~\cite{ifcnn}. This superior performance is evident in the enhanced clarity and naturalness of the fused images, which better handle the challenges of diverse exposure settings.

\begin{figure*}[!t]
\centering
\includegraphics[width=0.45\textwidth]{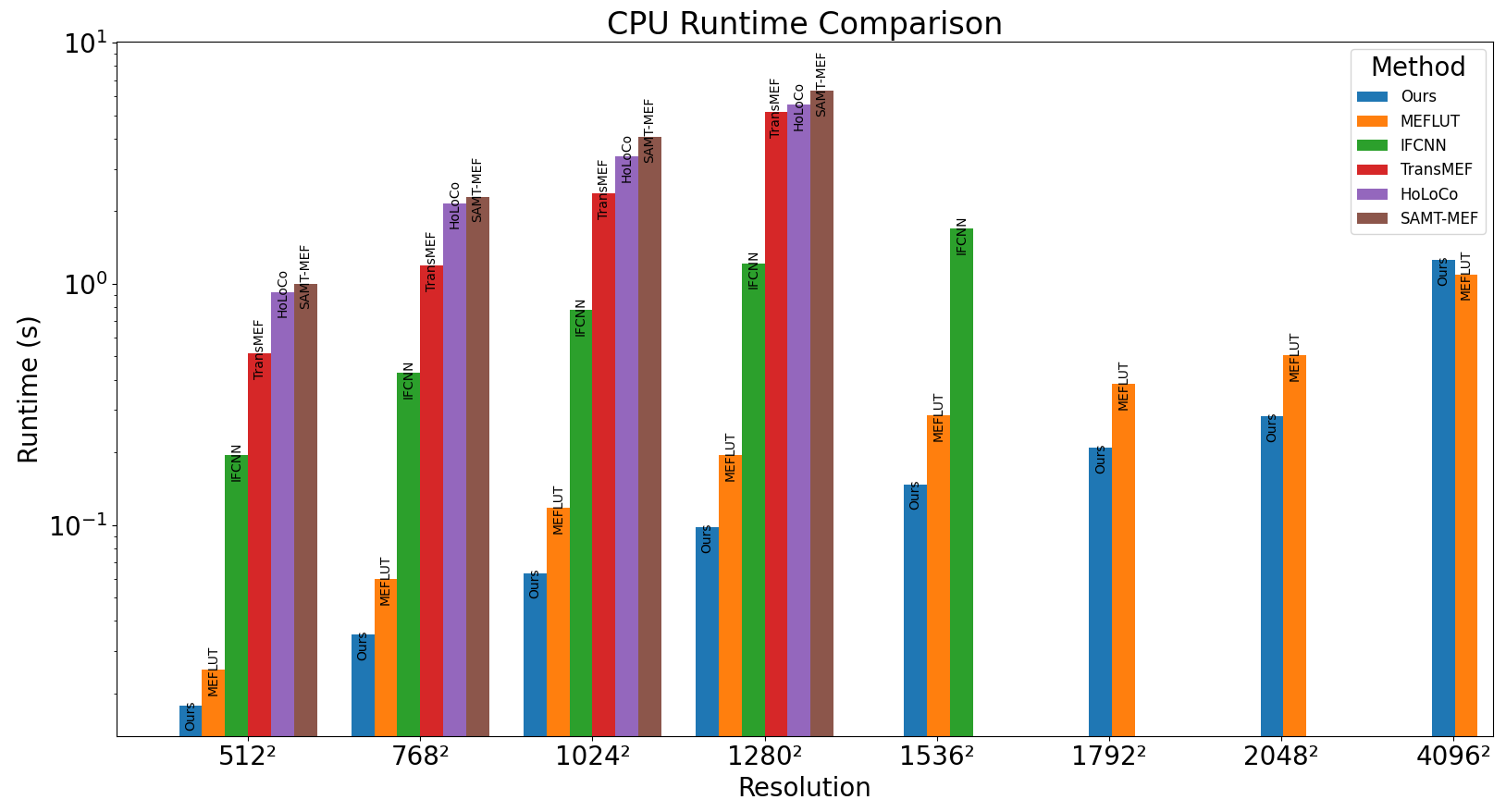}
~
\includegraphics[width=0.45\textwidth]{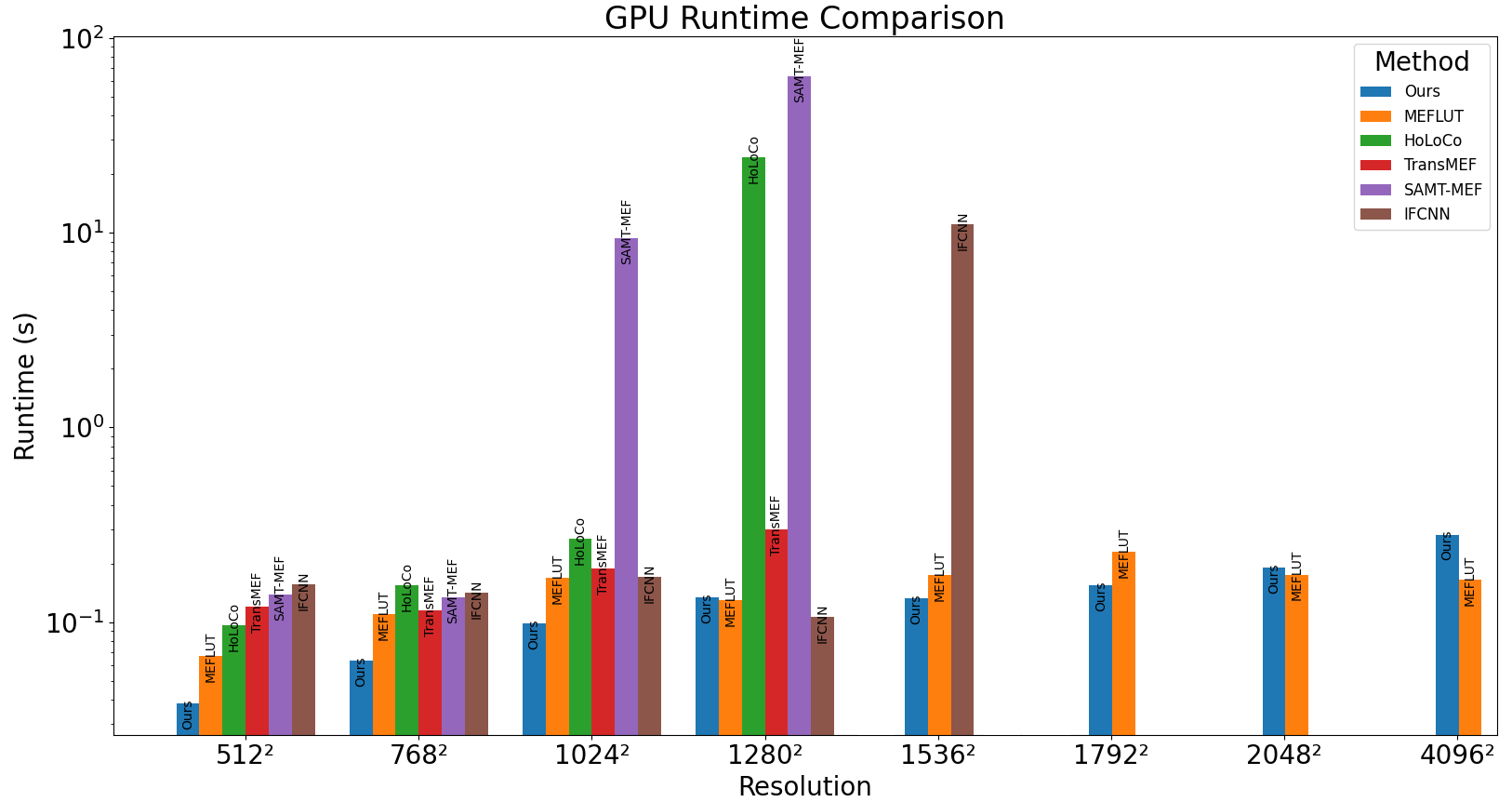}
\caption{Results for the runtime evaluations in CPU (left) and GPU (right).}
\label{fig:runtime}
\end{figure*}

\begin{figure}[!t]
\centering
\includegraphics[width=0.45\textwidth]{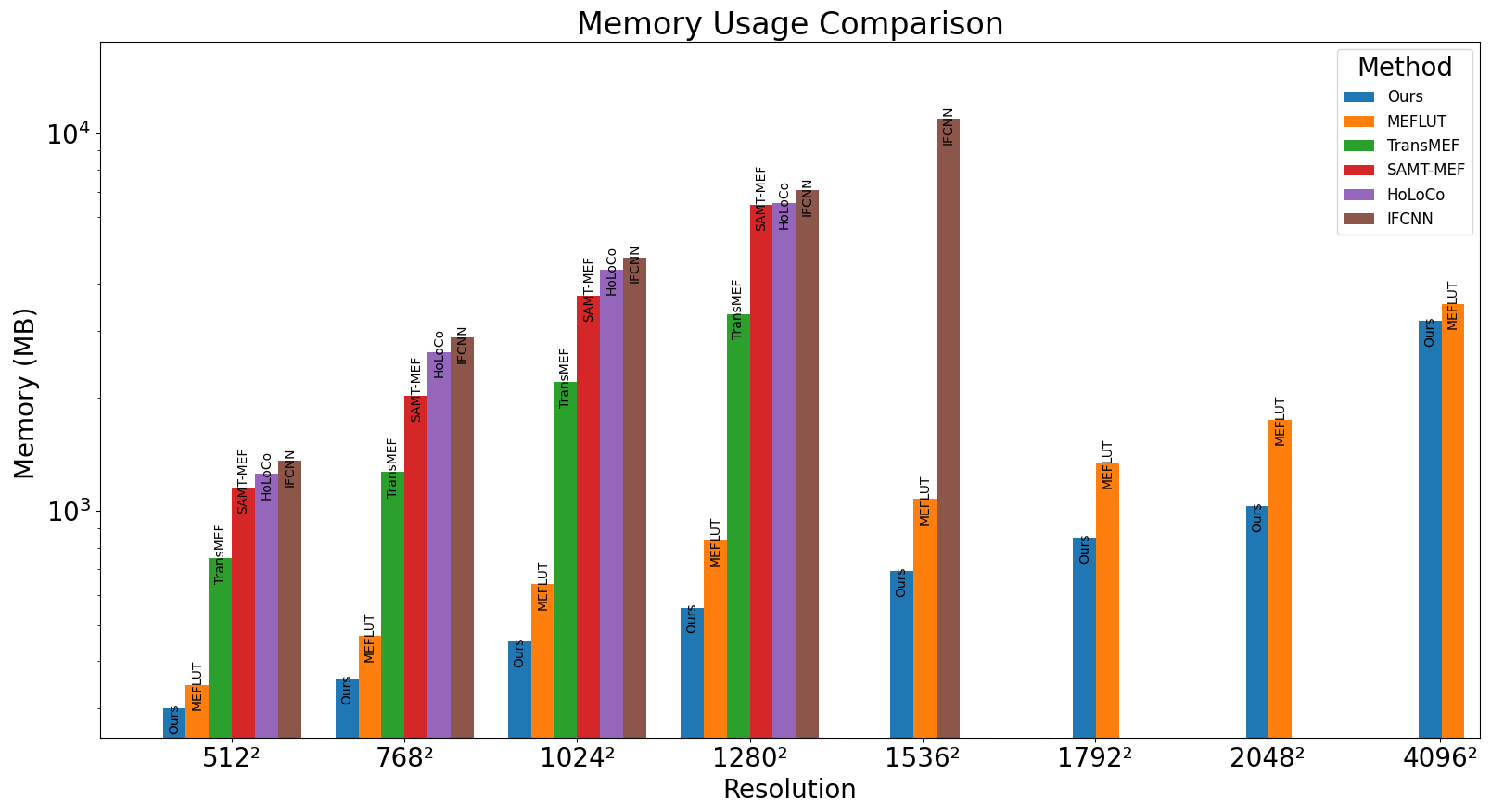}
\caption{Results for the memory evaluations.}
\label{fig:memory}
\end{figure}

The results for evaluating the computational resources are presented in Figs.~\ref{fig:runtime}~and~\ref{fig:memory} regarding runtime and memory, respectively. Our method achieves the lowest or near-lowest execution times across all resolutions and on both devices evaluation (CPU and GPU), even when compared to the MEFLUT~\cite{meflut} method that requires fewer operations (recall Tab.~\ref{tab:comp_results}). Specifically, our method is on average \textbf{1.67$\times$ faster on CPU} and \textbf{1.45$\times$ faster on GPU} compared to the fastest one on each resolution.
Regarding the memory usage evaluation, our method consistently outperforms all others, with the lowest memory consumption across all tested image resolutions. Specifically, our method is on average \textbf{1.41$\times$ more memory efficient} compared to the most efficient one on each resolution.

This makes our method effective in terms of the quality of fused images and highly efficient and scalable, ideal for real-time applications and high-resolution image processing, especially for applications with limited computational resources, such as smartphones. Hence, in comparison to other SOTA methods, our method provides a superior balance of computational efficiency and high-quality performance.

\subsection{Ablation studies}\label{sec:ablation}

In order to assess the effectiveness of our proposals, we conduct ablations studies related to (i) the neural network architecture (Sec.~\ref{sec:network_architecture} and \ref{sec:ssf_module}), (ii) the crop-like Gradient loss proposal (Sec.~\ref{sec:loss_function}), and (iii) the number of training epochs. For these studies, we trained our model with half input size resolution and using EV -1 and +1 as inputs, reduced the train epochs to 100 for studies (i) and (ii), and evaluated the model performance using only full-reference metrics, which we understand to be more reliable measures of performance when the ground-truth images are available. Specifically, we used SSIM, MS-SSIM, and PSNR, which are the most broadly used metrics for image quality assessment.

\textbf{Network Architecture.}
We present the results for the neural network architecture ablation in Tab.~\ref{tab:abblation_net}, where the ``Baseline'' method is the LPIENet~\cite{lpienet} architecture with modified inputs to agree with the MEF pipeline (i.e., allow multiple YUV frame inputs); the ``Filter Reduction (F.R.)'' column refers to diminishing the number of filters proposed in the original LPIENet model (they use 16 as the initial filter size for all channels, whereas we used 4 and 8 for the Y, and UV inputs, respectively); the ``Network Optimization (N.O.)'' column refers to the optimizations presented in Section~\ref{sec:network_architecture}; and the ``SSF-Module'' column refers to the usage of this new module as explained in Section~\ref{sec:ssf_module}. The ``Memory'' and ``Runtime'' were computed using a mobile device, as described in Section~\ref{sec:computation_eval}.

Our complete network yields the best quality measures for all the full-reference metrics. Moreover, the proposed network optimizations reduced the number of required operations and the amount of memory usage, with a slight increase in runtime compared to the Baseline model with reduced filters. Specifically related to the SSF module, note that it slightly impacts the computational efforts, with a small increase of 1.2\% of MACs, 5\% on memory usage, and 3.4\% on runtime. Nevertheless, such a small increase is justified due to the consistent improvement in all the quantitative metrics, and also supported by manual inspection of the results (i.e., qualitative analysis).

\begin{table*}[!t]
\centering
\caption{Ablation studies regarding the proposed network architecture: Filter Reduction (F.R. column), Network optimizations (N.O. column), and SSF module (SSF column). The MACs (G) column was computed using an image resolution of $4096^2$. The Memory and Runtime columns were computed using a mobile device. \textbf{Bold} values mark the best results, and {\ul underline} mark the second best ones.}
\resizebox{0.98\textwidth}{!}{
\begin{tabular}{l|ccc|ccc|ccc}
 &
  F.R. &
  N.O. &
  SSF &
  \textbf{MACs (G)} &
  \textbf{Memory (MB)} &
  \textbf{Runtime ($\mu$s)} &
  \textbf{MS-SSIM} &
  \textbf{SSIM} &
  \textbf{PSNR} \\ \hline
Baseline &
   &
   &
   &
  88.93 &
  1710.02 &
  5.63E+06 &
  {\ul 0.9087} &
  0.8532 &
  18.6070 \\
 &
  \checkmark &
   &
   &
  10.74 &
  {\ul 1125.29} &
  \textbf{1.49E+06} &
  0.9083 &
  {\ul 0.8612} &
  20.8147 \\
 &
  \checkmark &
  \checkmark &
   &
  \textbf{8.94} &
  \textbf{1100.66} &
  {\ul 1.72E+06} &
  0.9069 &
  0.8591 &
  {\ul 21.2006} \\ \hline
\textbf{MobileMEF} &
  \checkmark &
  \checkmark &
  \checkmark &
  {\ul 9.05} &
  1156.23 &
  1.82E+06 &
  \textbf{0.9088} &
  \textbf{0.8681} &
  \textbf{21.3015}
\end{tabular}
}
\label{tab:abblation_net}
\end{table*}

\textbf{Crop-like Gradient Loss.}
The crop-like Gradient loss ablation results are presented in Tab.~\ref{tab:abblation_loss}. We observe an increase in all three full-reference metrics. Specifically, the MS-SSIM improves by 2.5\%, the SSIM 3.8\%, and the PSNR 12.7\%.

\begin{table}[!t]
\centering
\caption{Ablation results for the proposed Gradient loss, $\mathcal{L}_{Grad}$.}
\begin{tabular}{l|ccc}
\textbf{$\mathcal{L}_{Grad}$} & \textbf{MS-SSIM} & \textbf{SSIM}   & \textbf{PSNR}    \\ \hline
Default  & 0.8867 & 0.8367 & 18.9027 \\
\textbf{Proposed} & \textbf{0.9088} & \textbf{0.8681} & \textbf{21.3015}
\end{tabular}
\label{tab:abblation_loss}
\end{table}

\textbf{Training Epochs.}
The results for the trained epochs ablation are presented in Tab.~\ref{tab:abblation_epochs}. We observe small differences among the model's performance on different epoch checkpoints, supported by the small standard deviation values of the metrics. This finding serves as an indicative of the model's robustness for overfitting.

\begin{table}[!t]
\centering
\caption{Epochs ablation results. \textbf{Bold} values mark the best results, and {\ul underline} mark the second best ones. The final rows correspond to the mean (MEAN) and standard deviation (STD) values, respectively.}
\begin{tabular}{l|ccc}
\textbf{Epochs} & \textbf{SSIM} & \textbf{MS-SSIM} & \textbf{PSNR} \\ \hline
100 & 0.8242 & 0.8931 & 20.5245 \\
200 & 0.8246 & 0.8926 & 20.5374 \\
\textbf{300} & \textbf{0.8270} & 0.8937 & {\ul 20.6532} \\
400 & 0.8251 & {\ul 0.8952} & \textbf{20.8056}\\
500 & {\ul 0.8266} & \textbf{0.8962} & 20.6528 \\ \hline
MEAN & 0.8255 & 0.8942 & 20.6347 \\
STD & 0.0013 & 0.0015 & 0.1135    
\end{tabular}
\label{tab:abblation_epochs}
\end{table}

\section{Conclusions}

In this work, we introduced MobileMEF, a novel MEF method based on an Encoder-Decoder deep learning architecture optimized for fast and efficient image processing on mobile devices. Our extensive evaluations demonstrate that our method outperforms SOTA approaches in most tested full-reference quality measures and computational efficiency (runtime and memory usage), making it highly suitable for real-time applications on devices with limited resources, such as smartphones.
Furthermore, our ablation studies confirmed the effectiveness of our architectural choices (i.e., the efficient building blocks and the SSF module), the crop-like Gradient loss proposal, and the robustness of our model across different training epochs.
Finally, our work offers a robust solution for MEF, balancing high image quality and low computational demands, thus paving the way for more efficient and effective HDR imaging on hardware-constrained applications. In future works, we intend to enhance the model's adaptability to a wider range of input conditions to mitigate its current limitations.

\bibliographystyle{IEEEtran}
\bibliography{references}

\end{document}